%
%

\input harvmac.tex
\noblackbox

\newcount\yearltd\yearltd=\year\advance\yearltd by 0

\input epsf
\newcount\figno
\figno=0
\def\fig#1#2#3{
\par\begingroup\parindent=0pt\leftskip=1cm\rightskip=1cm\parindent=0pt
\baselineskip=11pt
\global\advance\figno by 1
\midinsert
\epsfxsize=#3
\centerline{\epsfbox{#2}}
\vskip 12pt
{\bf Figure \the\figno:} #1\par
\endinsert\endgroup\par
}
\def\figlabel#1{\xdef#1{\the\figno}}


%
\def\np#1#2#3{Nucl. Phys. {\bf B#1} (#2) #3}
\def\plb#1#2#3{Phys. Lett. {\bf #1B} (#2) #3}

\def\physrev#1#2#3{Phys. Rev. {\bf D#1} (#2) #3}

\def\prep#1#2#3{Phys. Rep. {\bf #1} (#2) #3}

\def\atmp#1#2#3{Adv. Theor. Math. Phys. {\bf #1} (#2) #3}
\lref\juan{J. M. Maldacena, ``The large $N$ limit of superconformal field
theories and supergravity,'' hep-th/9711200, \atmp{2}{1998}{231}.}

\lref\gkp{S. S. Gubser, I. R. Klebanov and A. M. Polyakov,
``Gauge theory correlators
from non-critical string theory,'' hep-th/9802109,
\plb{428}{1998}{105}.}

\lref\witten{E. Witten, ``Anti-de-Sitter space and holography,''
hep-th/9802150, \atmp{2}{1998}{253}.}

\lref\wittenbar{E. Witten, ``Baryons in the $1/N$ expansion,''
Nucl. Phys. {\bf B160} (1979) 118.}

\lref\review{O. Aharony, S. S. Gubser, J. Maldacena, H. Ooguri and Y. Oz,
``Large $N$ field theories, string theory and gravity,'' hep-th/9905111,
\prep{323}{2000}{183}.}

\lref\klewit{I.~R.~Klebanov and E.~Witten,
``Superconformal field theory on threebranes at a Calabi-Yau  singularity,''
\np{536}{1998}{199}, hep-th/9807080.}

\lref\klegub{
S.~S.~Gubser and I.~R.~Klebanov,
``Baryons and domain walls in an $\cn = 1$ superconformal gauge theory,''
hep-th/9808075, \physrev{58}{1998}{125025}.}

\lref\morpless{
D.~R.~Morrison and M.~R.~Plesser,
``Non-spherical horizons. I,''
Adv.\ Theor.\ Math.\ Phys.\  {\bf 3}, 1 (1999),
hep-th/9810201.
}

\lref\GrisaruZN{
M.~T.~Grisaru, R.~C.~Myers and O.~Tafjord,
``SUSY and Goliath,''
JHEP {\bf 0008}, 040 (2000)
[arXiv:hep-th/0008015].
}

\lref\HashimotoZP{
A.~Hashimoto, S.~Hirano and N.~Itzhaki,
``Large branes in AdS and their field theory dual,''
JHEP {\bf 0008}, 051 (2000)
[arXiv:hep-th/0008016].
}

\lref\witson{ E.~Witten, ``Baryons and branes in anti de Sitter
space,'' JHEP {\bf 9807}, 006 (1998) [arXiv:hep-th/9805112].
}

\lref\abks{
O.~Aharony, M.~Berkooz, D.~Kutasov and N.~Seiberg,
``Linear dilatons, NS5-branes and holography,''
JHEP {\bf 9810}, 004 (1998)
[arXiv:hep-th/9808149].
}

\lref\giant{
J.~McGreevy, L.~Susskind and N.~Toumbas,
``Invasion of the giant gravitons from anti-de Sitter space,''
JHEP {\bf 0006}, 008 (2000)
[arXiv:hep-th/0003075].
}

\lref\bbns{
V.~Balasubramanian, M.~Berkooz, A.~Naqvi and M.~J.~Strassler,
``Giant gravitons in conformal field theory,''
JHEP {\bf 0204}, 034 (2002)
[arXiv:hep-th/0107119].
}

\lref\thooft{
G.~'t~Hooft,
``A planar diagram theory for strong interactions,''
Nucl. Phys. {\bf B72} (1974) 461.
}

\lref\corram{
S.~Corley, A.~Jevicki and S.~Ramgoolam,
``Exact correlators of giant gravitons from dual $N = 4$ SYM theory,''
Adv.\ Theor.\ Math.\ Phys.\  {\bf 5}, 808 (2001)
[arXiv:hep-th/0111222];
S.~Corley and S.~Ramgoolam,
``Finite factorization equations and sum rules for BPS correlators in
$N = 4$ SYM theory,''
Nucl.\ Phys.\ B {\bf 641}, 131 (2002)
[arXiv:hep-th/0205221].
}

\lref\cicuta{ G.~M.~Cicuta, ``Topological expansion for SO(N) And
Sp(2n) gauge theories,'' Lett.\ Nuovo Cim.\  {\bf 35}, 87 (1982).
}

\lref\bmn{ D.~Berenstein, J.~M.~Maldacena and H.Nastase,
``Strings in flat space and pp waves from N = 4 super Yang Mills,''
JHEP {\bf 0204}, 013 (2002)
[arXiv:hep-th/0202021].
}

\lref\gkbar{
S.~S.~Gubser and I.~R.~Klebanov,
``Baryons and domain walls in an $N = 1$ superconformal gauge theory,''
Phys.\ Rev.\ D {\bf 58}, 125025 (1998)
[arXiv:hep-th/9808075].
}

\lref\bhk{
D.~Berenstein, C.~P.~Herzog and I.~R.~Klebanov,
``Baryon spectra and AdS/CFT correspondence,''
JHEP {\bf 0206}, 047 (2002)
[arXiv:hep-th/0202150].
}

\lref\bhln{
V.~Balasubramanian, M.~X.~Huang, T.~S.~Levi and A.~Naqvi,
``Open strings from N = 4 super Yang-Mills,''
JHEP {\bf 0208}, 037 (2002)
[arXiv:hep-th/0204196].
}

\lref\BeasleyXV{
C.~E.~Beasley,
``BPS branes from baryons,''
arXiv:hep-th/0207125.
}

\lref\GukovKN{
S.~Gukov, M.~Rangamani and E.~Witten,
``Dibaryons, strings, and branes in AdS orbifold models,''
JHEP {\bf 9812}, 025 (1998)
[arXiv:hep-th/9811048].
}

\lref\MyersPS{
R.~C.~Myers,
``Dielectric-branes,''
JHEP {\bf 9912}, 022 (1999)
[arXiv:hep-th/9910053].
}

\def\cn{{\cal N}}
\def\CN{{\cal N}}
\def\IZ{\relax\ifmmode\hbox{Z\kern-.4em Z}\else{Z\kern-.4em Z}\fi}
\def\IR{\relax{\rm I\kern-.18em R}}

\def\frac#1#2{{#1 \over #2}}

\def\det{{\rm det}}
\def\tr{{\rm tr}}

\def\Pf{{\rm Pf}}

\Title{\vbox{\baselineskip12pt\hbox{hep-th/0211152}
\hbox{WIS/44/02-NOV-DPP}
}}
{\vbox{
{\centerline{`Holey Sheets' -- Pfaffians and Subdeterminants as}}
\vskip .1in
{\centerline{D-brane Operators in Large $N$ Gauge Theories}}
}}

\centerline{
Ofer Aharony$^{}$\foot{E-mail : {\tt Ofer.Aharony@weizmann.ac.il.}
Incumbent of the Joseph and Celia Reskin
career development chair.},
Yaron E. Antebi$^{}$\foot{E-mail : {\tt ayaron@weizmann.ac.il}.},
Micha Berkooz$^{}$\foot{E-mail : {\tt Micha.Berkooz@weizmann.ac.il.}
Incumbent of the Recanati career development chair of energy research.},
and Ram Fishman$^{}$\foot{E-mail : {\tt ramf@weizmann.ac.il}.}}
\bigskip
\centerline{{\it Dept. of Particle Physics,
The Weizmann Institute of Science, Rehovot 76100, Israel}}
\bigskip
\medskip
\noindent

In the AdS/CFT correspondence, wrapped D3-branes (such as ``giant
gravitons'') on the string theory side of the correspondence have been
identified with Pfaffian, determinant and subdeterminant operators on
the field theory side. We substantiate this identification by showing
that the presence of pairs of such operators in a correlation function
of a large $N$ gauge theory naturally leads to a modified 't Hooft
expansion including also worldsheets with boundaries. This happens
independently of supersymmetry or conformal invariance.

\Date{November 2002}

\newsec{Introduction and Summary of Results}

The correspondence \refs{\juan,\gkp,\witten,\review} between $AdS_5$
backgrounds of type IIB string theory and four dimensional gauge
theories provides the first explicit realization (above two
dimensions) of 't Hooft's general identification \thooft\ of large $N$
gauge theories with string theories, with the string coupling scaling
as $g_s \propto 1/N$.  The first examples of this correspondence, and
most of the cases that have been studied up to now, involved field
theories with fields only in adjoint or bifundamental
representations. Such field theories naturally map to closed string
theories \thooft. Backgrounds including D-branes filling $AdS_5$, or
filling $AdS_p$ subspaces of $AdS_5$, were also studied. In such
backgrounds there is also an open string sector on the string theory
side of the correspondence, corresponding to having additional fields
in the fundamental representation on the field theory side.

However, D-branes (and, therefore, an open string sector) can appear
also in the closed string theories corresponding to field theories
without any fundamental representation 
fields. In this paper we will be interested
only in localized D-branes, behaving like particles in anti-de Sitter
(AdS) space, which should correspond to local operators on the field
theory side of the correspondence. The first example of this was given
(and its field theory dual identified with a Pfaffian operator) in
\witson, and many other examples have been discovered since then.

A large class of examples of localized D-branes is related to ``giant
gravitons''. One of the most interesting results of the AdS/CFT
correspondence is a concrete realization of the general expectation that
high momentum excitations in gravity are described by large macroscopic
objects, rather than by short wavelength field excitations. It was
shown in \giant\ that gravitons with large angular momentum on the
$S^5$ become D3-branes wrapped on an $S^3$ inside the $S^5$ (similar
phenomena occur with M-branes in M-theory AdS backgrounds).
This example of Myers' dielectric effect \MyersPS\
goes under the name of ``giant gravitons'' \giant, and it is believed
to be part of a broader UV/IR mixing conspiracy in quantum gravity.
These states were conjectured in \bbns\ to correspond to subdeterminant
operators in the corresponding $\cn=4$ SYM theory. We will call the
field theory operators corresponding to D-brane states ``D-brane-type
operators''.

In this paper we attempt to understand how open strings can arise in a
theory involving only fields in the adjoint representation, whose
large $N$ expansion is naively given by a purely closed string
theory. We start by mapping string theory diagrams corresponding to
correlation functions of closed string vertex operators in the
presence of the D-brane to appropriately normalized correlation
functions on the field theory side (we could also similarly discuss
open string vertex operators, but we will not do this here).  We argue
that the consistency of the AdS/CFT correspondence requires that
appropriately normalized correlation functions involving pairs of
D-brane-type operators, such as Pfaffians and subdeterminants in the
$\cn=4$ SYM theory, should have an expansion in terms of open and
closed worldsheets\foot{An alternative way of seeing open strings in
the $\cn=4$ SYM theory, by analyzing small fluctuations around
D-brane-type operators in the BMN limit \bmn\ of the theory, was
discussed in \bhln.}. The bulk of the paper is devoted to showing that
such an expansion indeed exists and, moreover, that it exists for
Pfaffian and subdeterminant operators in any large $N$ theory, not
necessarily supersymmetric or conformal.

Our method involves developing the large $N$ expansion for
(appropriately normalized) correlation functions involving pairs of
Pfaffian or subdeterminant operators in $SO(2N)$ and $SU(N)$ theories,
respectively. Since these operators involve a number of fields of
order $N$, this expansion is quite different from the usual 't Hooft
expansion involving purely closed worldsheets (though, of course, the
``interior'' of Feynman diagrams, far away from the D-brane-type
operators, still maps to closed string worldsheets in the usual
way). In particular, we show that boundaries can arise in the
worldsheets of the Feynman diagrams, associated with external lines
emanating from the D-brane-type operators. We argue that we can map
any connected Feynman diagram to a worldsheet, which can have an
arbitrary number of boundaries and can also be disconnected.  Our main
result is the computation of the power of $N$ associated with these
Feynman diagrams in appropriately normalized correlation functions. We
show that it is precisely given (at leading order in $1/N$)
by $N^{\chi}$ where $\chi$ is the
Euler characteristic of the corresponding worldsheet, as expected for a string
theory involving open and closed strings whose string coupling scales
as $g_s \simeq 1/N$.  Our result depends on an assumption that we need
to make regarding the existence of a certain contour appearing in a
contour integration that arises in the correlation function
computation. We believe that this assumption is correct, but we have
not been able to prove it beyond the level of the free field
theory. It would be interesting to prove this assumption and to make
our results more rigorous. It would also be interesting to analyze the
$1/N$ corrections to our results.

Another prediction from string theory is that when we map a certain
connected diagram to a disconnected string theory worldsheet, its
value should factorize into the product of diagrams mapping separately
to each component of the worldsheet. We show that this
`factorization' property indeed holds, providing additional evidence for
our identification of the mapping of the Feynman diagrams to string
worldsheets.

The outline of the paper is the following. In section 2 we review some
known results regarding D-branes and ``giant gravitons'' in the
AdS/CFT correspondence. In section 3 we specify the class of field
theory quantities that we expect to map to correlation functions of
closed strings in the presence of a D-brane, and thus to have an
appropriate $1/N$ expansion. In section 4 we analyze the 't Hooft
large $N$ expansion for appropriate correlation functions involving
the Pfaffian operators in $SO(2N)$ theories. In section 5 we carry
over the analysis to the determinant and subdeterminant operators of
$SU(N)$ theories. We show that the appropriately normalized
correlation functions behave as expected for a theory involving both
closed and open strings. In all our field theory computations we do not use
supersymmetry, except in section 4.5 where we discuss special properties of
theories like
the $\cn=4$ SYM theory in which 2-point functions are not renormalized.

\newsec{D-branes in the AdS/CFT Correspondence}

The simplest operators to describe in the 't Hooft large $N$ limit
of gauge theories \thooft\ are gauge-invariant operators made of a
finite number of basic fields, which remains constant as $N \to
\infty$. An example of such operators is $\tr(X^l)$ if $X$ is a
field in the adjoint representation of the gauge group. From the
construction of the 't Hooft large $N$ limit it is clear that such
operators (if they are ``single-trace" operators, namely they
cannot be written as a product of two or more gauge-invariant
operators) map to local operators on the worldsheet of the string.

Indeed, this is exactly how things work in the AdS/CFT correspondence
\refs{\juan,\gkp,\witten,\review}, which, for instance, relates the
$SU(N)$ $\cn=4$
SYM theory to type IIB string theory on $AdS_5\times S^5$, and the
$SO(N)$ $\cn=4$ SYM theory to an orientifold of type IIB string
theory on $AdS_5\times RP^5$ \witson. Gauge-invariant operators in the
$\cn=4$ SYM theory involving $l \ll N$ basic fields, which are not a
product of smaller gauge-invariant operators, are mapped to integrated local
vertex operators on the string worldsheet. In particular, those
single-trace chiral
primary operators of the $\cn=4$ SYM theory which are made of a small number of
basic fields (or, equivalently, the ones in small representations of
the $SU(4)_R$ global symmetry group) are mapped to type IIB supergravity
fields.

The description of such operators as local operators on the string
worldsheet is valid only for $l \ll N$. For instance, the operator
$\tr(X^{N+1})$ in an $SU(N)$ gauge theory may be written as a linear
combination of products of lower order traces, so it should not
correspond to an independent vertex operator in the theory, but it is
not known how to see this directly in string theory\foot{Presumably
this is related to our poor understanding of string theory in RR
backgrounds; the analogous bound is understood in cases which only
involve NS-NS backgrounds, as in \abks.}. The usual 't Hooft large $N$
expansion seems to break down when used for operators involving a
large number of fundamental fields (although techniques exist for some
baryonic operators, as in \wittenbar), and apriori it is not clear that
their correlation functions should have any reasonable large $N$
expansion.  In particular, trace-type operators $\tr(X^l)$ with large
$l$ mix significantly with multi-trace operators\foot{A basis of
operators which have diagonal 2-point functions in free field theory
was constructed in \corram. For $l \simeq \sqrt{N}$ this mixing plays
an important role in the recent study of string theory in plane wave
backgrounds.}.

Some particular operators involving a large number of basic fields have
been identified with D-branes in the type IIB background which are
localized in $AdS_5$ (we will only discuss localized D-branes in this
paper).
The D-brane states that were matched with the field theory side fall
into two classes. One class includes D-branes which are topologically
stable on the string theory side. This means that there is
some charge in the theory and that the D-brane state is the lightest
state with this charge, ensuring its stability. One can then usually
identify the operator in the field theory side by looking for
the operator of lowest dimension with the same charge (in
supersymmetric theories these usually turn out to be chiral operators in
short representations of the superconformal algebra, which restricts
the operator mixing and simplifies the identification).
Another class includes D-brane states which are dynamically
stable, such as ``giant gravitons''. One is
typically less sure about the identification of these states in the
field theory. One of the goals of this paper is to support existing
proposals for the identification of the dynamically stable branes.

{\bf Topologically stable branes}

The first identification of a D-brane with a field theory operator was
presented in \witson. In that case the identification 
involved the
Pfaffian operator of $\cn=4$ supersymmetric $SO(2N)$ gauge theories, defined by
\eqn\Pfaffian{
\Pf(X) = \frac1{(2N)!}  \epsilon^{i_1i_2\cdots
i_{2N}}X_{i_1i_2}X_{i_3i_4}\cdots X_{i_{2N-1}i_{2N}}, }
where the $X$'s
can be any of the six adjoint scalar fields in the theory, contracted in
their $SO(6)_R$ indices in a symmetric traceless manner so as to form
a chiral primary operator. This was identified with a D3-brane
wrapped on the non-trivial 3-cycle in the orientifold $RP^5$. This
identification was motivated by the fact that this brane carries a
$\IZ_2$ charge (coming from the fact that the relevant cohomology of
$RP^5$ is $\IZ_2$) which may be identified with the $\IZ_2$ charge in
the center of the $SO(2N)$ gauge theory, under which the Pfaffian
operator is charged (but operators including less $X$'s are not). It
was supported by various computations, including the mass of this
D3-brane (related to the conformal dimension of the corresponding
operator), its $SU(4)_R$ transformation properties, etc.

Similar operators, corresponding to D3-branes wrapped on non-trivial
3-cycles, were
found to exist in other $AdS_5$ backgrounds as well. In $AdS_5\times
M$ backgrounds
corresponding to product gauge groups with bi-fundamental matter
fields, it was found \refs{\gkbar,\morpless,\GukovKN,
\bhk,\BeasleyXV} that D3-branes wrapped on
3-cycles of $M$ could be identified with dibaryon-type operators made
of $N$ bi-fundamental fields. We will not discuss this case here,
but we expect it to behave similarly to the Pfaffian case which we will discuss
in detail.

{\bf Dynamically stable branes}

An example of dynamically stable (supersymmetric) branes in
$AdS_5\times S^5$ was given in \giant. In this case they are D3-branes
wrapping a 3-sphere in $S^5$, which is topologically trivial, and
the branes are dynamically stabilized by their angular momentum (which
includes contributions both from orbital angular momentum and from
interactions with background fields). The size of these branes grows as
their angular momentum increases, reaching the maximum possible size
for an $S^3$ in $S^5$ when the angular momentum reaches the maximum
possible value of $N$ (for a single-trace operator).  It was argued in
\giant\ that it is in terms of these branes, rather than fundamental
strings, that one should describe Kaluza-Klein gravitons of large
angular momentum on $S^5$, and they are therefore called ``giant
gravitons''.

In \bbns\ it was suggested that these D-branes should be identified
with determinant and subdeterminant operators of the $\cn=4$ SYM
theory, and that the same identification applies also to
similar theories with lower
supersymmetry (although the details may vary from case to case). For
the $\cn=4$ theory, the giant gravitons with angular momentum $L \leq
N$ ($N-L \ll N$) were identified with the subdeterminant operators
\eqn\Sdetop{\det_L(X)=\frac1{L!(N-L)!}  \epsilon_{i_1\cdots
i_Li_{L+1}\cdots i_N} \epsilon^{j_1\cdots j_Li_{L+1}\cdots i_{N}}
X_{j_{1}}^{i_{1}}\cdots X_{j_{L}}^{i_{L}},}
where $X$ is one of the
complex scalar fields of the theory.  The subdeterminant operators
with $L=2,3,\cdots,N$ form an alternative basis to the algebra of
gauge-invariant operators,
instead of the ``standard'' $\tr(X^l)$ basis.

Some of the reasons for this identification are \bbns:
\item{1.} Some correlation functions of these operators are protected
and hence can be computed in the free theory. The results can be used
to show that these operators form a better basis in the sense that
their mixing is smaller by powers of $N$. The mixing was analyzed more
generally in \corram, supporting this conclusion.
\item{2.} In some cases, several topologically stable
 branes can be combined to
give a dynamically stable brane (for example a pair of Pfaffian
operators in the $SO(2N)$ theory), facilitating the identification of
the latter.
\item{3.} In some cases one can compute the expectation value of these
operators along flat directions.
\item{4.} This identification naturally explains the bound $L \leq N$
on the angular momentum of the ``giant gravitons'' coming from
D3-branes wrapped on an $S^3$ in $S^5$.

In this paper we will support this identification further for the
determinant and subdeterminant operators in $SU(N)$ gauge theory. We expect
that a similar analysis will hold for other cases with product gauge
groups and adjoint or bi-fundamental matter.

{\bf Summary}

In summary, D3-branes wrapped on the compact part of space-time in the
AdS/CFT correspondence seem to be generally related to gauge-invariant
operators formed by multiplying a large number $l \simeq N$ of basic
fields in the theory with anti-symmetric contractions. We will analyze the
implications of this for the field theory in the next section,
focusing on the simplest case of the $\cn=4$ SYM theory with gauge
groups $SO(2N)$ and $SU(N)$. We will discuss in this paper only
D-branes which are completely wrapped on the compact space, so as to
give particles on AdS space (which are mapped to local operators in
the field theory)\foot{D-branes which are not localized but go all
the way to the boundary of $AdS_5$ simply give rise to matter fields
in the fundamental representation in the field theory.}.
There are also various other possible localized D-brane
configurations, such as ``giant gravitons" extended in $AdS_5$
\refs{\GrisaruZN,\HashimotoZP}\foot{A
conjecture on the identification of these branes in the
field theory was presented in \corram.} or D-branes
which must have fundamental strings ending on them \witson, and we will not
discuss these here, though the analysis of the next section should
apply also to them.

\newsec{Open String Diagrams in Field Theory}

The correlation functions of single-trace operators involving a small
number of basic fields in the $\cn=4$ SYM theory map (in the string
perturbation theory approximation) to correlation functions of the
appropriate closed string vertex operators (for chiral primary
operators these can be computed in the supergravity approximation for
large $g_{YM}^2 N$). As discussed above, some operators ${\cal O}_D$
in the SYM theory are believed to map to D-brane states. So, their
correlation functions should correspond to processes involving
D-branes. General processes involving D-brane creation and annihilation
cannot be studied in string perturbation theory. However, other processes
involving D-branes, such as scattering of closed strings off D-branes, can
be studied in string perturbation theory, and we will focus on these
processes in this paper.

Since we are interested in processes which do not create or annihilate
D-branes, we need to have a D-brane present in the initial and final states
(for simplicity, we will discuss here the case of a single D-brane). It is
simplest to discuss such processes in global AdS space-time, which
maps via the AdS/CFT correspondence to the $\cn=4$ SYM theory on $S^3\times
\IR$. Of course, correlation functions on $S^3\times \IR$ are related to
correlation functions on $\IR^4$ by the usual conformal transformation
involved in radial quantization. A process involving a D-brane in the
initial state maps via this relation to an insertion of the
D-brane-type operator ${\cal O}_D$ at the origin of $\IR^4$, while
having the same D-brane in the final state maps to an insertion of
${\cal O}_D^{\dagger}$ at infinity. Of course, we could also have
slightly different initial and final D-brane states if we put in
different operators at zero and infinity.

The simplest processes we can discuss in string perturbation theory are
those involving some number of closed string vertex operators in the
presence of the D-brane (namely, allowing string worldsheets which have
boundaries
on the D-brane). The discussion above suggests that if the closed string
vertex operators correspond to field theory
operators ${\cal O}_i(x_i)$, then such a
correlation function in closed+open string theory should be related to
the field theory correlation function
\eqn\fieldcorr{\langle {\cal O}_D(0) {\cal O}_D^{\dagger}
(\infty) \prod_i {\cal O}_i(x_i) \rangle.}

What is the exact relation between these correlation
functions\foot{See \bbns\ and \corram\ for previous discussions of such
correlation functions.} ? In the AdS/CFT correspondence, correlation
functions of local operators ${\cal O}(x)$ in the CFT are related to the string
theory partition function by:
\eqn\correspondence{\vev{e^{\int d^4 x \phi_0 (x)
{\cal O}(x)}}_{CFT}=Z_{string}[\phi (x,z)_{z=0} \simeq \phi_ 0 (x)].}
Here, $\phi_0(x)$ is an arbitrary function, and the field $\phi(x,z)$
is the field in AdS space
corresponding to the operator ${\cal O}(x)$. Correlation
functions $\vev{{\cal O}(x_1)...{\cal O}(x_k)}$ in field theory are the
coefficients of $\phi_0(x_1)...\phi_0(x_k)$ in the expansion of the
left hand side.  The right hand side is the full partition function of
string theory with the boundary condition that the field $\phi$ has
the value $\phi_0$ (up to an appropriate power of the radial
coordinate $z$)
on the boundary of AdS. To get the coefficient of
$\phi_0(x_1)...\phi_0(x_k)$ on the right hand side one sums over all
closed string topologies with $k$ insertions of the integrated
vertex operators of the
field $\phi$ at the appropriate points $x_i$.

The expansion of $\vev{{\cal O}(x_1)...{\cal O}(x_k)}$ in field theory
Feynman diagrams will include disconnected vacuum diagrams. To get rid
of them one normalizes by the value of the left hand side for
$\phi_0=0$, which is just the vacuum partition function. This
normalization removes the disconnected closed surfaces with no vertex
operator insertions from the string theory expansion of the right hand
side. Similarly, to compute \fieldcorr\ we will use string theory in a
background with a D-brane, and the expansion on the string theory side
will include also disconnected surfaces with boundary with no vertex
operator insertions, which we would like to remove in the
normalization.  And, in the field theory side, when computing the
generating function $\langle e^{\int d^4 x
\phi_ 0 (x){\cal O}(x)} {\cal O}_D(0) {\cal O}_D^{\dagger}(\infty)
\rangle_{CFT}$
it is natural to again normalize by the value
of this expression for $\phi_0=0$, dividing by
$\langle {\cal
O}_D(0) {\cal O}_D^{\dagger}(\infty) \rangle$.

Thus, we would like to suggest that
the relation between correlators of D-brane-type operators in the
field theory and computations in a closed+open string theory with
boundaries on the corresponding D-brane takes the form
\eqn\postulate{{{\langle e^{\int d^4 x
\phi_ 0 (x){\cal O}(x)} {\cal O}_D(0) {\cal O}_D^{\dagger}(\infty)
\rangle_{CFT}}\over {\langle {\cal O}_D(0) {\cal O}_D^{\dagger}(\infty)}
\rangle_{CFT}}
=Z_{open+closed,D-Brane}[\phi (x,z)_{z=0} \simeq \phi_ 0 (x)],}
with no vacuum diagrams appearing on the right-hand side.
This means that
the normalized correlation functions
\eqn\ratio{{{\langle {\cal O}_D(0)
{\cal O}_D^{\dagger}(\infty) \prod_i {\cal O}_i(x_i) \rangle}\over
{\langle {\cal O}_D(0) {\cal O}_D^{\dagger} (\infty) \rangle}}}
should have a large $N$ expansion involving surfaces with or without
boundaries, all containing vertex operator insertions.
Note that in field theory it
is easy to compute the denominator of \ratio. The D-brane-type operators
diagonalize the 2-point functions and have a fixed dimension
$\Delta_D$, so it is simply given by $1 / (\infty)^{2 \Delta_D}$.
This is an infinite factor that will appear also in the numerator of \ratio,
so the ratio \ratio\ should be finite.

We suggest that finite ratios like \ratio\ should
map to the correlation function of the closed string vertex operators
corresponding to the operators ${\cal O}_i$ in the presence of the
D-brane. This suggestion agrees with the leading term in the large $N$
limit of \ratio, which comes from taking a disconnected correlation
function $\langle {\cal O}_D(0) {\cal O}_D^{\dagger}(\infty) \rangle
\langle \prod_i {\cal O}_i(x_i) \rangle$ in the numerator, and looking
only at planar diagram contributions to $\vev{\prod_i {\cal O}_i(x_i)}$; this
obviously agrees with the leading term in $g_s$ on the right hand
side of \postulate,
arising from a sphere diagram which does not note the presence
of the D-brane. Note that the correspondence works in the simplest way
when we normalize the
operators ${\cal O}_i$ such that their 2-point functions scale as
$N^2$ in the large $N$ limit, so we
will use this normalization throughout this paper (in the
normalization we will use, an example of such operators is
$N {\rm tr}(X^l)$). In this normalization diagrams whose topology has
Euler characteristic $\chi$ scale as $N^{\chi}$ in the large $N$ limit.

In string theory, the right-hand side of \postulate\ has
an expansion in powers of $g_s$ involving even and odd powers of
$g_s$, coming from all diagrams (connected and disconnected, with or
without boundaries) appearing in string perturbation theory.
Using \postulate, the AdS/CFT correspondence maps this
to having a similar good $1/N$ expansion for
\ratio\ in the $\cn=4$ SYM theory.  The identification of the D-brane-type
operators is only well-understood in
the weakly curved string theory, corresponding to $g_{YM}^2 N \gg
1$. However, we hope that the same properties should hold for any
value of $g_{YM}^2 N$, so they should be visible also in perturbation
theory (in the 't Hooft limit). In the remainder of the paper we will
show (with a mild assumption) that this is in fact true in any large $N$
gauge theory (not necessarily having a weakly curved string theory dual),
though we used supersymmetric theories to motivate it.
Note that even though we used conformal invariance of the field theory
in the discussion above to motivate the expression \ratio, conformal
invariance does not seem to influence the general properties of the
large $N$ expansion (though it certainly affects the space-time dependence
of correlation functions), so we could have a similar expansion for
\ratio\ even in non-conformal theories (with arbitrary positions for
the D-brane-type operators), and we will show that this is
indeed the case.

\newsec{$SO(2N)$ Gauge Theories}

In this section we will derive the large $N$ topological expansion for
\ratio\ in $SO(2N)$ gauge theories with Pfaffian operators.
In these theories, \ratio\ takes the form
\eqn\norgencorr{\frac{\langle\Pf(X)\Pf(X^{\dagger})
N\tr(X^{J_1})N\tr(X^{J_2})...N\tr({X^{\dagger}}^{L_1})
N\tr({X^{\dagger}}^{L_2})...\rangle}
{\langle\Pf(X)\Pf(X^{\dagger})\rangle},}
choosing a particularly simple form of ${\cal O}_i$ (which should not
affect the results) and suppressing the space-time positions of the operators.
As discussed above, based on string theory we expect such a large $N$ expansion
to exist, and to include all topologies of compact surfaces with
boundaries\foot{Closed surfaces will also appear, but they describe
processes in which there is no interaction between the closed strings
and the D-brane.}, at least for the $\CN=4$ SYM theory.

The outline of this section is the following. In \S4.1 we will review
the known large $N$ expansion for correlators of single trace
operators. In \S4.2 we will analyze the form and $N$-dependence of a
general diagram in the correlator $\langle\Pf(X)\Pf(X^{\dagger})\rangle$,
and in \S4.3
we will extend the analysis to diagrams in correlators which include
trace operators as well. We will identify the topology corresponding to
each diagram and find diagrams with all topologies with boundaries. In
\S4.4 we will argue that \norgencorr\ indeed has a good
large $N$ expansion, and
that the contribution of diagrams which have a specific topology
(as defined in \S4.2 and
\S4.3) is proportional as expected to $N^{\chi}$, where $\chi$ is that
topology's Euler characteristic. The analysis will apply to any $SO(2N)$
gauge theory, not just supersymmetric ones. In \S4.5 we will discuss
some special features of the $\CN=4$ SYM theory.

\subsec{Review and definitions}

We are interested in general $SO(2N)$ gauge theories with complex
scalar fields $X$ in
the adjoint representation (it is easy to generalize our results also
to real scalar fields, or to fields of higher spin,
including the gauge bosons).
The scalar fields are $2N\times 2N$ complex antisymmetric matrices.
For simplicity we
will focus on operators involving only a single complex scalar field
$X$.
The gauge invariant operators
\eqn\Trop{\tr(X^L)=
X_{i_1i_2}X_{i_2i_3}X_{i_3i_4}\cdots X_{i_Li_1},} for $L$ even and
independent of $N$ in the large $N$ limit, correspond in the 't
Hooft large $N$ expansion to closed string vertex operators (in
the AdS/CFT correspondence they map to supergravity states with
angular momentum $L$ on $RP^5$). The Pfaffian operator is defined
as:
\eqn\Pfaffiann{ \Pf(X) = \frac1{(2N)!}
\epsilon^{i_1i_2\cdots i_{2N}}X_{i_1i_2}X_{i_3i_4}\cdots
X_{i_{2N-1}i_{2N}}. }

The 't Hooft large $N$ expansion (in which $\lambda=2Ng_{YM}^2$ is
held fixed when $N\to\infty$) of correlators with only single trace
operators is well known \refs{\thooft,\cicuta},
and we will just review it briefly
here. Feynman diagrams can be written in the
double line (``fatgraph'') notation, in which the propagator $\langle
X_{ij}X^{\dagger}_{kl}\rangle $ is represented by two lines, each
having one fundamental index. The propagator is given by:
\eqn\SOprop{ \vev{X_{ij}(x_1)X^{\dagger}_{kl}(x_2)} \propto
{{\delta_{il}\delta_{jk}-\delta_{ik}\delta_{jl}}\over {(x_1-x_2)^2}}. }
We see that the indices are the same on both ends of the
propagator but the existence of the two terms means that the
direction of the lines in the Feynman diagram is not preserved.
We will discuss theories (like supersymmetric Yang-Mills theories)
whose interaction vertices
are given by single traces, and there is always one
index running on each closed loop. We will normalize
the fields (obviously,
the normalization does not affect normalized correlators) such that
the whole action is proportional to $1 / g_{YM}^2 = 2N / \lambda$, and then
each propagator carries a factor of $1/2N$ and each interaction vertex
a factor of $2N$ (ignoring the $\lambda$-dependence).
Interpreting the Feynman diagram as a triangulation of a two
dimensional surface, it then follows that the expansion in Feynman
diagrams is an expansion in two dimensional compact closed topologies (both
oriented and unoriented), where Feynman diagrams with
each topology carry a power of
$2N$ equal to their Euler characteristic (times some function of $\lambda$).

\subsec{The $\langle\Pf(X)\Pf(X^{\dagger})\rangle$ correlation function}

In this section we will analyze the form and the $N$-dependence of
Feynman diagrams in the $\langle\Pf(X)\Pf(X^{\dagger})\rangle$ correlation
function.

\fig{The free diagram in the $\langle\Pf(X)\Pf(X^{\dagger})\rangle$
correlation function. Each Pfaffian is made of $N$ basic fields.
The propagators are unoriented, and could involve an exchange of the
two lines.
}{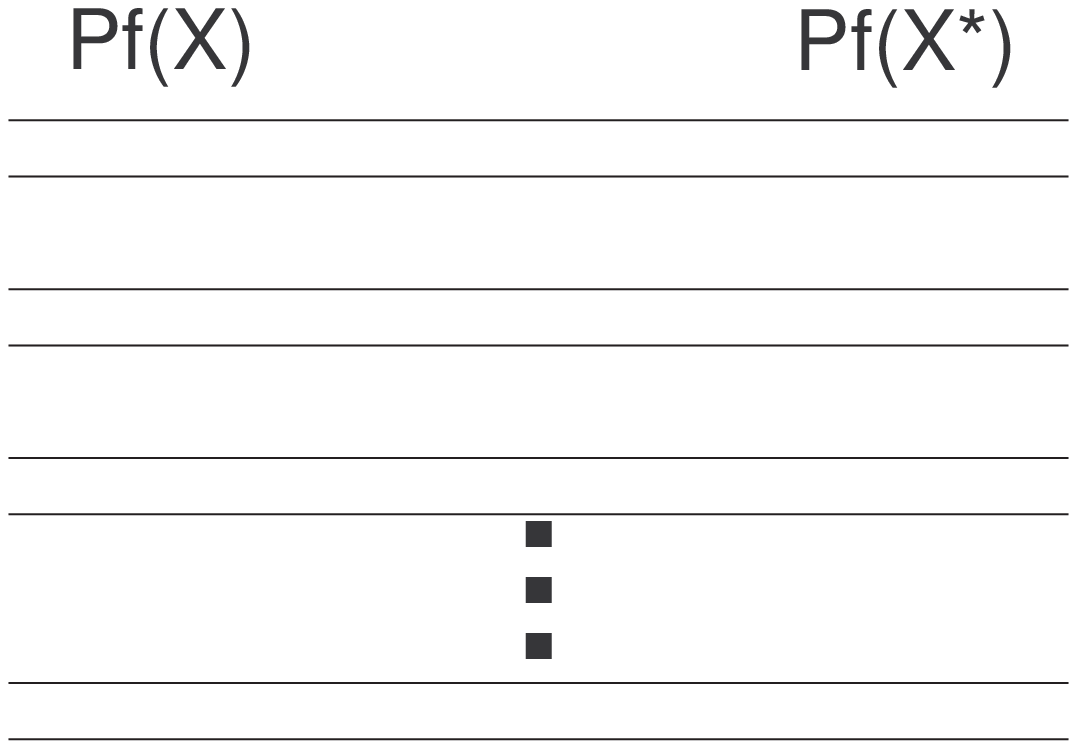}{2.3truein} \figlabel{\freepf}

We will refer to $\Pf(X)$ as the ``incoming Pfaffian'' and to
$\Pf(X^{\dagger})$ as the ``outgoing Pfaffian''. The simplest diagram is the
free diagram shown in figure \freepf, and it will be convenient to normalize
other diagrams by dividing by it. In the $\CN=4$
SYM theory this is in fact the only contribution to
$\langle\Pf(X)\Pf(X^{\dagger})\rangle$, because the two-point function of
chiral primary operators is not renormalized. However, our
discussion here (until section 4.5)
will apply also to non-supersymmetric field theories.

\fig{A general
diagram for the $\langle\Pf(X)\Pf(X^{\dagger})\rangle$ correlation
function. A connected diagram may have several connected components
after removing the Pfaffian vertices.}{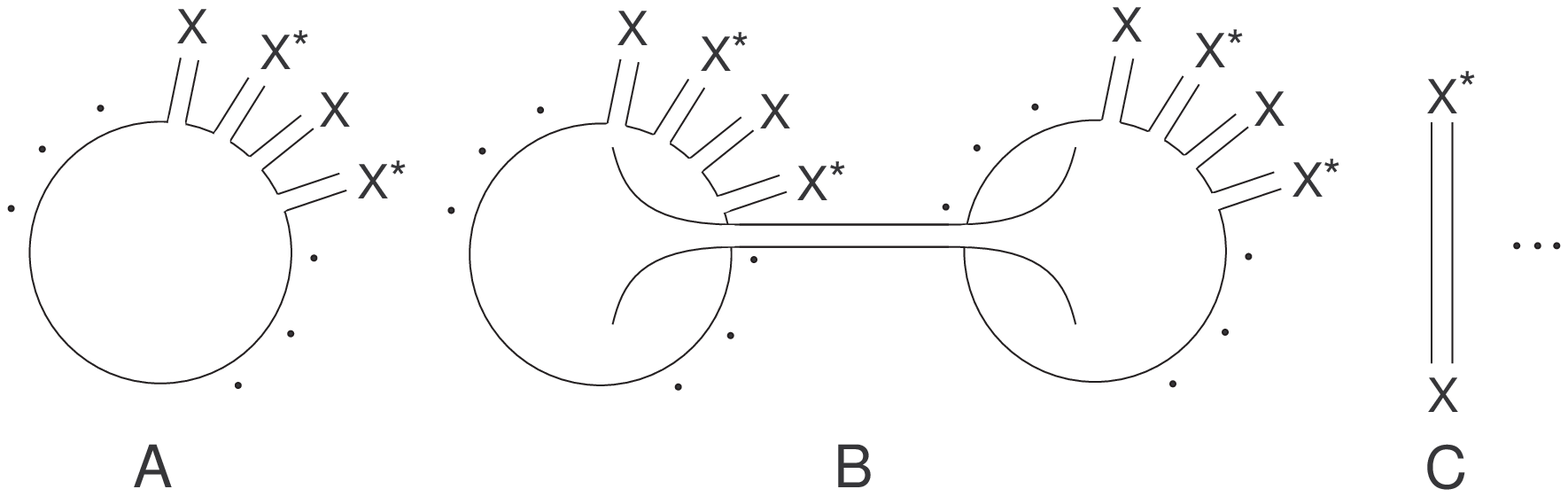}{5truein}
\figlabel{\generalpf}

The general form of a Feynman diagram for this correlation function is
shown in figure \generalpf. There are $N$ external legs originating from each
of the Pfaffians, which are arranged in several connected components
in the diagram\foot{All the diagrams we discuss here will be
connected, but we will separate the diagrams into components which are
connected or disconnected when we remove the Pfaffian vertices. All
external legs from each Pfaffian come from the same spacetime point,
but it is more instructive to ignore this fact in the drawings.}. The
simplest connected component is just a single propagator between an
$X$ and an $X^{\dagger}$ and we will refer to it occasionally as a ``trivial
component''. Each single line from an external leg must connect to
another external line. However, in the $SO(2N)$ theory (unlike the
$SU(N)$ case which will be discussed in the next section) no two indices
on the incoming (or outgoing) Pfaffian can be equal, so every single
line coming from an external leg must go between the two
Pfaffians. This implies that the external legs are arranged in each of
the connected components in an alternating order, as in figure \generalpf, and
that each component must include an equal number of legs from the
incoming and outgoing Pfaffians.

\fig{Amputating the diagram results in a vacuum diagram. A dashed
line represents a boundary. (a) A disk diagram. (b) A M\" obius
strip. (c) A cylinder. }{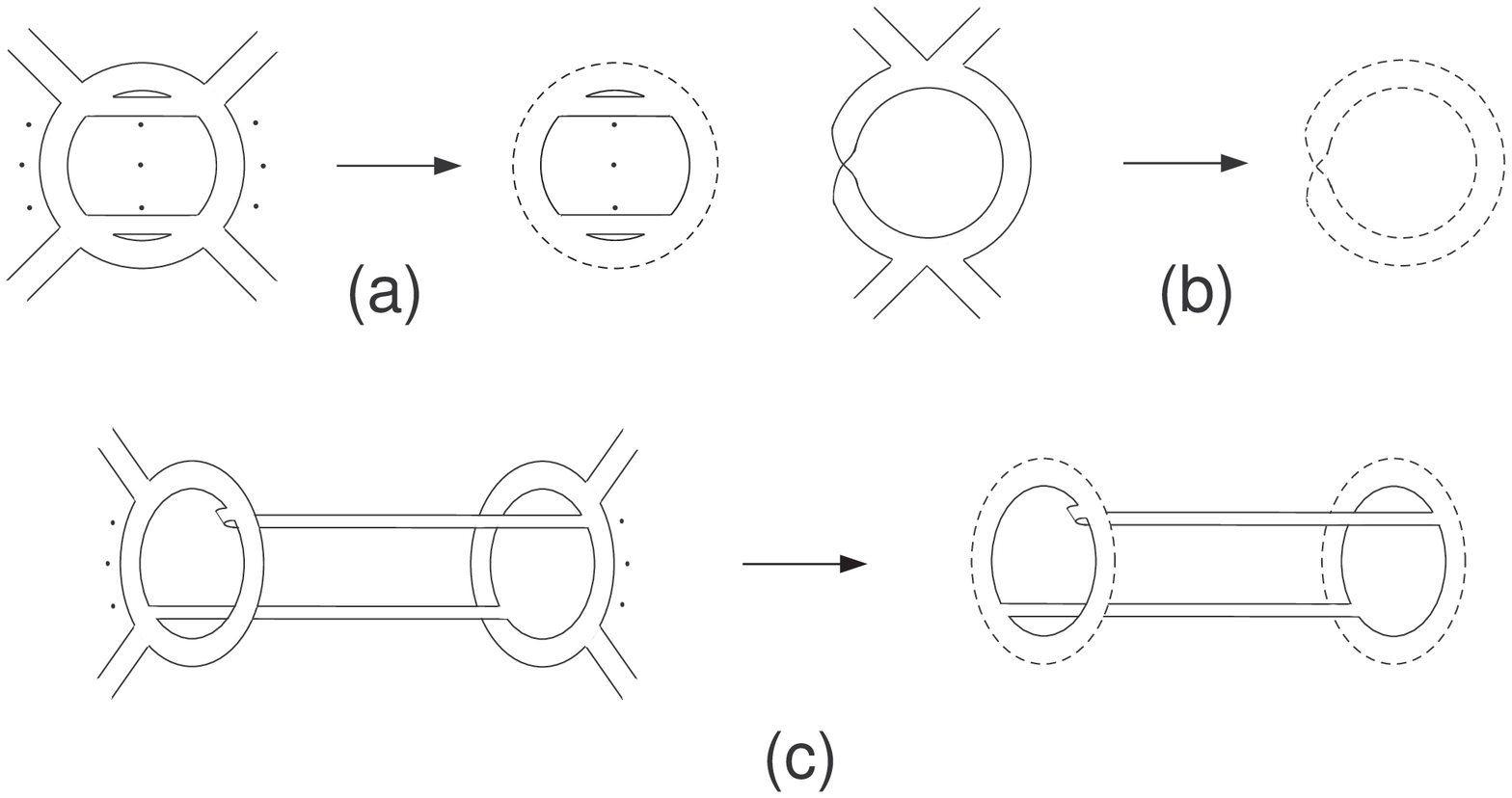}{5.2truein}
\figlabel{\amputation}

Suppose that we amputate the lines coming from external legs in any
Feynman diagram (as shown in figure \amputation). The remaining legs,
vertices and
faces form a vacuum diagram, of the form considered in
\refs{\thooft,\cicuta}, in which each
amputated line
becomes (part of) a boundary of the surface\foot{We will not
really be relating these amputated components
to vacuum diagrams, since
these are different diagrams with different numerical values,
but this is a convenient way of thinking about the
$N$-dependence.}. Each component of the diagram gives rise in this way
to a (not necessarily oriented)
compact surface. The amputation of a trivial
component will be taken to be a disk.  For instance, in the example of
figure \generalpf, the connected component A has the topology of a disk, and B
has the topology of an annulus (we can create higher genus surfaces by
adding handles and crosscaps). We will call these bounded surfaces the
internal surfaces of the components.

This construction is motivated by the $N$-dependence of the diagrams,
which is :

\medskip

\item{1.}  {\bf External legs}: each external leg contributes a factor of
$\frac1{2N}$ from its propagator, so the $2N$ external legs give
$(2N)^{-2N}$.  Note that although in every trivial component the two
external legs are really one, our convention for the internal topology
of such a component (a disk) will make up for the missing factor of $2N$
(see item 2).

\medskip

\item{2.} {\bf The topology of the internal surfaces}: this
contribution is calculated as in the usual large $N$ expansion
\refs{\thooft,\cicuta}. The contribution of an internal surface is
$(2N)^\chi$, where $\chi$ is the Euler characteristic of the
internal surface, including the contribution of the boundaries.
Here we do not sum over the external indices running on
the boundaries of the surface (they do not contribute $N$
factors). We will deal with these next.

\medskip

\item{3.} {\bf Combinatoric factors}: the number of possible ways of
connecting external indices to the internal diagrams. This comes from
the possible indices of the external legs, the number of
equivalent diagrams arising from permutations of the external legs, and
possible $N$-dependent symmetry factors.

\medskip

The general picture that we will develop is that the internal topologies
as they appear in these diagrams become the topologies of
string theory. To do this we will need to verify that the topologies
come with the correct power of $1/N$, and we will do so in \S4.4 (at
leading order in $1/N$).

To determine the $N$-dependence, we need to evaluate the contribution
of all three sources above. The first two are straightforward but the
third will require some consideration. As a warm-up exercise, we will
begin with the $N$ counting of the free diagram, shown in
figure \freepf, considering each of the three sources as follows:

\item{1.} The external leg propagators give $(2N)^{-2N}$.

\item{2.} The $N$ trivial components count as $N$ disks with 
$\chi=1$, so we get $(2N)^{N}$.

\item{3.} We will compute the combinatoric factor in a way that is
cumbersome for the free
diagram, but will be more useful later. First we have two
factors of $\frac1{(2N)!}$ from the definition of the Pfaffian
\Pfaffian. Both the incoming and outgoing Pfaffian are made out of $N$
fields in a symmetric combination, hence there are $(N!)^2$ ways of
choosing the order of the $X$'s and the $X^{\dagger}$'s in figure \freepf.
Since choosing the same pairs in a different order results in the same
diagram, we counted every diagram $N!$ times, so we should divide by
this symmetry factor.  For a given pair we can still swap the two
lines of the $X$ and the two lines of the $X^{\dagger}$. This contributes a
factor of $4$ for every propagator. Again we counted each diagram
several times so we should divide by a symmetry factor. If we swap the
lines at both the incoming and outgoing ends of a propagator, we get
the same diagram.
The symmetry factor
here is $2$ for each propagator.
After we chose how to connect the incoming fields to the
outgoing fields, we sum over all possible arrangements of indices in
the diagram. This gives us a factor of $(2N)!$ since we need to give
each of the $2N$ external lines a different color index.

The $N$-dependence of the free diagram is therefore:
\eqn\FreeDiag{ \rm{free\ diagram} \propto
\frac{(2N)!}{((2N)!)^2} \frac{(2^NN!)^2}{2^NN!}(2N)^{-N} =
\frac{N!}{(2N)!}N^{-N}. }

Proceeding to the general diagram, we consider a diagram with some
number of connected components with different topologies, as in
figure \generalpf. The contribution of the three sources is:

\item{1.} Again, this is just $(2N)^{-2N}$.

\item{2.} From the internal topologies we get the factor $(2N)^{\sum\chi}$
where $\sum\chi$ is the sum of the Euler characteristics of all
the internal surfaces (including the trivial disks).

\item{3.} We have, as before, two factors of $\frac1{(2N)!}$ coming
from the definition \Pfaffian. Now, we can change the order of all the
$N$ legs of each of the two Pfaffians, and we can swap the two
lines of each external leg since there is no orientation for the
lines. These operations will not change the value of the diagram and
they add a factor of $(2^NN!)^2$. Since all external lines start on
the left and reach the right we have $(2N)!$ different ways for
assigning values to the indices of the external legs. However, as for
the free diagram, some diagrams were overcounted when we calculated
the combinatoric factor. This happens when the diagram has some
symmetry. 
This can happen in two ways.
First, whenever the
diagram has several identical components, some of the diagrams
obtained by reordering the external legs give the same diagram but
were counted as different diagrams. Each set of $k$ identical components
gives a symmetry factor of $k!$. Second, in every component
which has just two external legs (one incoming and one outgoing),
swapping the external legs of both gives back the same diagram, so
there's a symmetry factor of $2^F$, for a diagram with $F$
two-legged components\foot{Additional, accidental symmetries may be
present when the internal components are made of special, symmetric
diagrams, and should be corrected for. These are rare, however, when
we go to large 't Hooft coupling and the diagrams become dense, so we
will ignore them.}.

Putting all the factors together we get:
\eqn\generalNonorm{
\frac{(2^NN!)^2}{\rm{Symmetry\ Factors}}
\frac{(2N)!}{((2N)!)^2}(2N)^{-2N}(2N)^{\sum\chi} ,}
where the symmetry factors include the two contributions described above.
After
normalizing (for convenience) by dividing by the free diagram we find that the
$N$-dependence of a general diagram is :
\eqn\generalSON{
\frac{N!}{N^N} (2N)^{\sum\chi} \frac1{\rm{Symmetry\ Factors}}.}
Note that the above analysis made use only of the color structure
of the theory, and not of any special features of the theory,
such as supersymmetry.

\subsec{Correlation functions with traces}

\fig{A typical diagram for
$\langle\Pf(X)\Pf(X^{\dagger})\tr(X^J)\tr({X^{\dagger}}^J)\rangle$.
The `x' stands for
insertions. When there is a $U(1)$ symmetry acting on $X$, as in SYM theory,
the two insertions must be in the same component, but this need not be
the case in general.}
{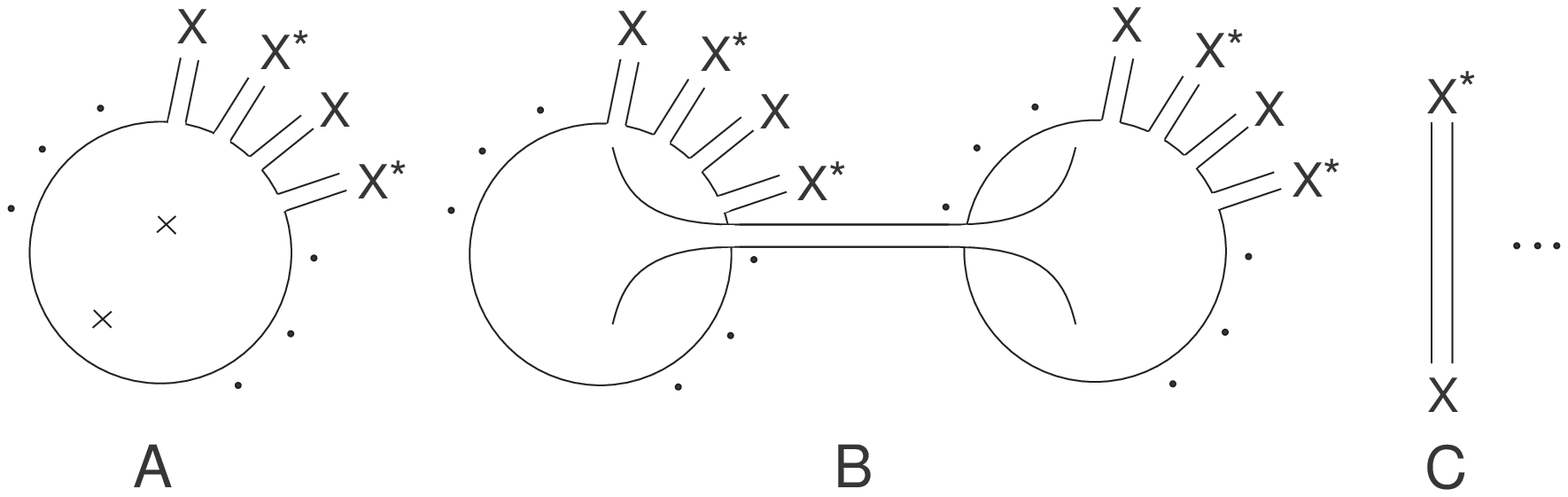}{5truein}
\figlabel{\generalpftr}

Consider now correlators of the form
\eqn\gencorr{\langle\Pf(X)\Pf(X^{\dagger})
N\tr(X^{J_1})N\tr(X^{J_2})...N\tr({X^{\dagger}}^{L_1})
N\tr({X^{\dagger}}^{L_2})...\rangle,}
which in string theory are related to scattering of type IIB
supergravity particles off the D-brane. The general field theory
diagram for this correlator will include some components
which include trace insertions (in the interior of the surface as
usual) and some that do not (figure \generalpftr\ shows an
example in which only one
component has insertions, but of course this need not be the case 
in general).

The insertions function as ordinary interaction 
vertices in the theory as far as
$N$-dependence is concerned (when we take each trace to come with a
factor of $N$ as above). The $N$ dependence of these diagrams is
therefore identical to that of the same diagram in
$\langle\Pf(X)\Pf(X^{\dagger})\rangle$,
given (after normalization) by \generalSON.

\subsec{Expansion in topologies}

In this section we will derive the large $N$ expansion of the normalized
correlators \norgencorr.
As we discussed in section 3, in the expansion of the numerator of
\norgencorr\ on the string theory side, each surface in which the vertex
operators are inserted appears together with all possible additions of
disconnected empty (no vertex operators) surfaces. 
In the ratio \norgencorr, however,
the normalization cancels the
contribution of all `empty' surfaces, and one is left with an expansion in
$N^{\chi}$ where $\chi$ is the Euler characteristic of
the topology of the surfaces in
which the vertex operators are
inserted.

We saw that we can characterize the field theory diagrams by the
topologies of the internal components, and that a subset of these
components, which we will call the interacting component, contains the
trace insertions (this terminology is somewhat misleading, since the
interacting `component' may consist of several disconnected
components).  It is natural to identify the topology of the
interacting component with the corresponding topology in the string
theory expansion of \norgencorr\ (again, these may be
disconnected). There are many diagrams with the same topology
for the interacting component: different number of external legs can
be attached to it, and there may be other internal components with
arbitrary topologies. For example, the disk amplitude in string theory
will correspond to all diagrams for which there is a single
interacting component with the topology of a disk.

In order to justify this identification we need to check that
the (normalized) contribution of all diagrams whose interacting
component has Euler characteristic $\chi$ is proportional to $N^{\chi}$,
as it is on the string theory side.

The perturbative expansion of
string theory amplitudes exhibits also `factorization' properties. For example,
the `two disks'
contribution to the scattering of two closed strings
off the D-brane (each disk here contains two vertex operator
insertions, one for an incoming and one for an outgoing string) is
just the product of the two separate `one disk' contributions to the
scattering of each closed string off the D-brane `on its own'. If our
identification is correct, we expect to see the same behavior on the
field theory side, though it is far from obvious why this should be
the case since we are not discussing disconnected diagrams on the
field theory side.

We will start by writing down explicit expressions for both the
numerator and the denominator of \norgencorr\ as sums of diagrams. We will
then evaluate these expressions using the saddle-point method, and get
a result of the expected form 
for \norgencorr, their ratio. Finally, we will use
this result to confirm the `factorization' property explained above.

\medskip
\noindent {{\it 4.4.1} \bf $\langle\Pf(X)\Pf(X^{\dagger})\rangle$}
\medskip

Let $D_{k,\chi}$ denote the sum of all connected diagrams (with no
vertex operator insertions) with $k$ incoming and $k$ outgoing
legs (arranged in an alternating order) such that their internal
topology has Euler characteristic equal to $\chi$, without
including the $N$-dependent contributions to the diagrams (but
including the space-time dependence which we suppress). 
For $k=1$ we will divide
$D_{1,\chi}$ by two, to simplify the notation below.

Now, consider the sum of all diagrams which have $L_{k,\chi}$ connected
components with $k$ incoming (and $k$ outgoing) external legs and internal
Euler characteristic $\chi$. Normalized by the free diagram, it is given by:
\eqn\onediagram{\frac{N!}{N^N}\prod_{k,\chi}\frac{(2N)^{\chi
L_{k,\chi}}D_{k,\chi}^{L_{k,\chi}}}{L_{k,\chi}!}.}
The product extends over $k=1,\cdots,N$, $\chi=1,0,-1,\cdots$. It is easy to
see that the factorial terms in the denominators give the right
symmetry factors for each diagram in the sum. Recall that there is another
symmetry factor of $2^{-L_{1,\chi}}$, which we take into account by 
our insertion of a $\frac{1}{2}$ in the definition of $D_{1,\chi}$.

To obtain the full correlation function $\vev{\Pf(X) \Pf(X^{\dagger})}$ we 
need to sum over all possible diagrams, which is the same as summing
over all possibilities for $L_{k,\chi}$, and we get:
\eqn\alldiagrams{\sum_{L_{k,\chi} \geq 0}\frac{N!}{N^N}\prod_{k,\chi}
\frac{(2N)^{\chi L_{k,\chi}}D_{k,\chi}^{L_{k,\chi}}}{L_{k,\chi}!},}
where the sum is constrained by the condition $\sum_{k,\chi}k
L_{k,\chi}=N$. We will denote the same sum \alldiagrams\ with the constraint
$\sum_{k,\chi} k L_{k,\chi} = n$ by $a_n$.

\medskip
\noindent {{\it 4.4.2 $\langle\Pf(X)\Pf(X^{\dagger})
\tr(X^{J_1})\tr(X^{J_2})\cdots\tr({X^{\dagger}}^{L_1})
\tr({X^{\dagger}}^{L_2})\cdots\rangle$}}
\medskip

Turning our attention to correlation functions with insertions, let us
evaluate the contribution of all diagrams whose interacting component has Euler
characteristic $\chi_0$ \foot{Note that the
interacting `component' may actually consist of several
disconnected components, and then $\chi_0$ could be as large as the
number of disconnected components. 
}.
Every such diagram is characterized by the number $k_0$ of
incoming (and outgoing) external legs connected to its interacting
component, $1\leq k_0\leq N$, and, as before, by $L_{k,\chi}$, the number
of (insertion free) components with Euler characteristic $\chi$
and $k$ incoming
(and outgoing) external legs. These satisfy $\sum_{k,\chi}k
L_{k,\chi}+k_0=N$.

Let $C_{k,\chi}$ denote the sum of all diagrams with $k$ incoming (and $k$
outgoing) legs (arranged in alternate order on each component) which
include the
trace insertions and such that their internal topology has Euler
characteristic $\chi$, again without including the $N$-dependence of
the diagrams. We could also consider here just a specific topology
with this Euler characteristic, and this will be useful in \S4.4.4,
but it wouldn't affect the qualitative result.

The contribution of all diagrams that have specific values of $k_0$
and $L_{k,\chi}$ is
\eqn\onediagramtrace{\frac{N!}{N^N}(2N)^{\chi_0}C_{k_0,\chi_0}
\prod_{k,\chi}\frac{(2N)^{\chi
L_{k,\chi}}D_{k,\chi}^{L_{k,\chi}}}{L_{k,\chi}!}.}
The total correlation function comes from summing over all values of $k_0$ and
$L_{k,\chi}$:
\eqn\alldiagramstrace{\sum_{L_{k,\chi};k_0>0}\frac{N!}{N^N}(2N)^{\chi_0}
C_{k_0,\chi_0}\prod_{k,\chi}\frac{(2N)^{\chi
L_{k,\chi}}D_{k,\chi}^{L_{k,\chi}}}{L_{k,\chi}!},}
where the sum is constrained by the condition $\sum_{k,\chi}k
L_{k,\chi}+k_0=N$. We will denote the same sum \alldiagramstrace\ 
with the constraint
$\sum_{k,\chi}k L_{k,\chi}+k_0=n$ by $b_n$.

\medskip
\noindent {{\it 4.4.3 Saddle point evaluation}}
\medskip

We want to show that the leading $N$-dependence of the 
contribution of diagrams whose
interacting component has Euler characteristic $\chi_0$ to \norgencorr,
which is the ratio of
\alldiagramstrace\ and \alldiagrams, is given by
$(2N)^{\chi_0}$. We will evaluate it using the saddle point method, as
follows.

Define the `partition functions':
\eqn\partition{\eqalign{ Z_1(\beta)
&\equiv
\sum_{n=0}^{\infty} a_n \beta^n = \cr
&= \frac{N!}{N^N}\sum_{\left\{L_{k,\chi}\geq0\right\}}
\beta^{\sum_{k\geq1,\chi\leq1}kL_{k,\chi}}
\prod_{k,\chi}\frac{(2N)^{\chi
L_{k,\chi}}D_{k,\chi}^{L_{k,\chi}}}{L_{k,\chi}!}= \cr
&=\frac{N!}{N^N}\prod_{k,\chi}\sum_{\left\{L_{k,\chi}\geq0\right\}}
\frac{(\beta^k
D_{k,\chi}(2N)^{\chi})^{L_{k,\chi}}}{L_{k,\chi}!}
=\frac{N!}{N^N}e^{h(\beta)},}}
where $h(\beta)\equiv \sum_{k\geq1,\chi\leq1}
\beta^k D_{k,\chi}(2N)^{\chi}$, and
\eqn\partitiontrace{\eqalign{ Z_2(\beta) &\equiv
\sum_{n=0}^{\infty} b_n \beta^n = \cr
&= \frac{N!}{N^N}(2N)^{\chi_0}\sum_{\left\{L_{k,\chi}\geq0,k_0 > 0\right\}}
\beta^{k_0+\sum_{k\geq1,\chi\leq1}kL_{k,\chi}}
C_{k_0,\chi_0}\prod_{k,\chi}\frac{(2N)^{\chi
L_{k,\chi}}D_{k,\chi}^{L_{k,\chi}}}{L_{k,\chi}!}= \cr
&=(2N)^{\chi_0}(\sum_{k_0\geq1}\beta^{k_0}
C_{k_0,\chi_0})\frac{N!}{N^N}\prod_{k,\chi}
\sum_{\left\{L_{k,\chi}\geq0\right\}}\frac{(\beta^k
D_{k,\chi}(2N)^{\chi})^{L_{k,\chi}}}{L_{k,\chi}!} = \cr
&=(2N)^{\chi_0}g_{\chi_0}(\beta)Z_1(\beta),}}
where $g_{\chi_0}(\beta)\equiv \sum_{k\geq1}\beta^k C_{k,\chi_0}$.
We are interested in computing the correlation functions
$\alldiagrams=a_N$ and
$\alldiagramstrace=b_N$. These coefficients are given by the contour
integrals:
\eqn\integralN{\eqalign{ a_N &=\frac{1}{2\pi i} \oint
Z_1(\beta)\beta^{-(N+1)}d\beta, \cr
b_N &=\frac{1}{2\pi i}
\oint Z_2(\beta)\beta^{-(N+1)}d\beta=(2N)^{\chi_0}\frac{1}{2\pi i}
\oint g_{\chi_0}(\beta)Z_1(\beta)\beta^{-(N+1)}d\beta.} }

We will begin by analyzing $a_N$.
Keeping only the two leading (first and zeroth) powers
in $N$ in $h(\beta)$ (the leading term cannot vanish since the theory
has a propagator) we get \eqn\integralN {a_N \simeq
\frac{1}{2\pi i} \frac{N!}{N^N}\oint
e^{2Nf(\beta)}e^{f_0(\beta)}\frac{d\beta}{\beta},}
where we define $f(\beta)\equiv\sum_{k\geq1}\beta^k D_{k,1}
- \frac{1}{2}\log(\beta)$, including the leading terms in $h$,
and $f_0(\beta)\equiv\sum_{k\geq1}\beta^k
D_{k,0}$, with the subleading terms;
both functions are independent of $N$.

Since $N\gg 1$, we may use the saddle point method. In this method, one
considers a contour along which the imaginary part of $f(\beta)$ is
fixed (this is the trajectory of steepest descent for the real part of
$f$). Such a contour, if it exists, will pass through some extremal
points of $f(\beta)$ where $Re(f(\beta))$ is maximal. Since $N$ is
large, the main contribution to the integral comes from
the vicinity of those points, where we can approximate the integrand as a
(very narrow) Gaussian. The question is whether in fact there is such
a contour with constant $Im(f(\beta))$ 
which circles the origin, possibly going to
infinity as long as $Re(f(\beta))\rightarrow -\infty$ there.  We will
conjecture that this is indeed the case.

Assume first, for simplicity, that there is one saddle
point $\beta_0$, for which $f^\prime (\beta_0)=0$. We then get
\eqn\saddlepoint{a_N\simeq \frac{1}{2\pi i}
\frac{N!}{N^N}e^{i\alpha}\frac{\sqrt{2\pi}e^{2N|f(\beta_0)|}
e^{f_0(\beta_0)}}{\sqrt{2N
|f^{\prime \prime}(\beta_0)|}\beta_0}}
where $e^{i\alpha}$ is a phase determined by the direction of steepest
descent.

As the simplest example, consider a theory in which only
$D_{1,1}\neq0$. The only diagram is the free one so we expect to get
$a_N=(2D_{1,1})^N$ (recall that our expressions for the diagrams are
normalized by the free diagram with a propagator equal to one and that
$D_{1,1}$
was defined as half the propagator). Indeed, in this case we have
$f(\beta)=D_{1,1}\beta-\frac{1}{2}\log(\beta)$ so $\beta_0=1/2D_{1,1}$
and so \eqn\free{a_N\simeq \frac{1}{2\pi i} \frac{N!}{N^N}
e^{i\alpha}\frac{\sqrt{2\pi}e^{2N(\frac{1}{2}+\frac{1}{2}
\log(2D_{1,1}))}}{\sqrt{2N 2D_{1,1}^2}(2D_{1,1})^{-1}}=
-i e^{i\alpha}(2D_{1,1})^N
\frac{N! e^N}{N^N \sqrt{2\pi N}}=(2D_{1,1})^N }
where in the last equality we used the Stirling
approximation.
In this example we can explicitly verify that a contour of `steepest descent'
encircling the origin in fact exists, justifying the saddle point evaluation,
and that the phases cancel out properly.

We can make a similar analysis for $b_N$. The only difference is the
appearance of another ($N$-independent) function $g_{\chi_0}(\beta)$ in the
integrand alongside the steep exponential. The saddle point method
applies here as well, and since the gaussian is very narrow, one can take
$g_{\chi_0}(\beta)$ as a constant equal to $g_{\chi_0}(\beta_0)$.
Aside from this
factor, the result is the same (since the saddle point is the same),
so we get:
\eqn\saddlepointtrace{b_N\simeq \frac{1}{2\pi i}
\frac{N!}{N^N}(2N)^{\chi_0}e^{i\alpha}\frac{\sqrt{2\pi}g_{\chi_0}(\beta_0)
e^{2N|f(\beta_0)|}e^{f_0(\beta_0)}}{\sqrt{2N|f^{\prime
\prime}(\beta_0)|}
\beta_0}.}
Most terms cancel out when we take the ratio
\eqn\ratiotwo{\frac{b_N}{a_N}\simeq g_{\chi_0}(\beta_0)(2N)^{\chi_0},}
which is indeed proportional to $(2N)^{\chi_0}$ as expected.

When there is more than one saddle point along the integration contour 
things don't cancel out so
neatly. However, if there is one saddle point for which $Re(f(\beta))$
is the largest, its contribution to the integral strongly dominates in
both $a_N$ and $b_N$, so we again get the same result. We run into
trouble when there are two or more saddle points for which
$Re(f(\beta))$ is the same. $Re(f(\beta))$ is a function of $\beta$
with power series coefficients that depend on the 't Hooft coupling
$\lambda$. It would be surprising if this function would have the
property that it has the same value at two extremal points for all
values of $\lambda$, so we may expect our result \ratiotwo\ 
to hold at least for
generic values of the 't Hooft coupling.

To summarize, up to an assumption about the existence of an appropriate
contour of constant phase encircling the origin, we have shown that the
correlation functions behave as we claimed, with the leading contribution
of diagrams whose interacting component has Euler characteristic
$\chi_0$ coming with a coefficient of $N^{\chi_0}$ as in string theory.

\medskip
\noindent {{\it 4.4.4  Factorization}}
\medskip

There is another test we can make of our identification. In
string theory, the scattering of two string states
$\vev{\Pf(X)\Pf(X^{\dagger})N\tr(X^l)N\tr(X^{\dagger
l})N\tr(X^m)N\tr(X^{\dagger m})}$ gets a contribution from the
disconnected topology having two components, with Euler
characteristics $\chi_1$ and $\chi_2$, the first of which contains
the $\{\tr(X^l), \tr(X^{\dagger l})\}$ insertions and the second of
which contains the $\{\tr(X^m), \tr(X^{\dagger m})\}$ insertions.
This contribution is the product of the $\chi_1$ contribution to
$\vev{\Pf(X)\Pf(X^{\dagger})N\tr(X^l)N\tr(X^{\dagger l})}$ and the
$\chi_2$ contribution to
$\vev{\Pf(X)\Pf(X^{\dagger})N\tr(X^m)N\tr(X^{\dagger m})}$ (all 
divided by $\vev{\Pf(X)\Pf(X^{\dagger})}$).

The same equality should hold in the field theory. Note that in
section 4.4.2 we did not need to assume that the interacting component
was connected. According to
\ratiotwo, we should check that
\eqn\factorization{ g_{\chi_1\otimes
\chi_2}(\beta_0)(2N)^{\chi_1+\chi_2}=g_{\chi_1}(\beta_0)g_{\chi_2}(\beta_0)
(2N)^{\chi_1}(2N)^{\chi_2}.}
But $g_{\chi}(\beta)=\sum_{k\geq1}\beta^k C_{k,\chi}$. Recall that
$C_{k,\chi_1\otimes \chi_2}$ is the sum of all diagrams
(without the $N$-dependence)
with $k$ external legs and internal topology
consisting of two components, one with topology $\chi_1$ and the other
$\chi_2$. The $k$ external legs will be divided between these two
components, so $C_{k,\chi_1\otimes \chi_2}=\sum_{k_1+k_2=k}
C_{k_1,\chi_1}C_{k_2,\chi_2}$. It then follows that $\sum_{k_1,k_2}
\beta^{k_1}\beta^{k_2}C_{k_1,\chi_1}C_{k_2,\chi_2}=\sum_k \beta^k
C_{k,\chi_1\otimes \chi_2}$ or in other words, that we have $g_{\chi_1\otimes
\chi_2}(\beta)=g_{\chi_1}(\beta)g_{\chi_2}(\beta)$, which
confirms \factorization.

It is important to note that the above analysis is applicable to any
theory with the same `color structure'. We made no use of the special
properties (most notably, the supersymmetry) of the $\CN=4$ SYM theory,
which has a known equivalent string theory in which we know
that the Pfaffian operator should actually correspond to a D-brane.

\subsec{The supersymmetric theory}

In this section we discuss the $\CN=4$ SYM theory with $SO(2N)$ gauge
group. This theory contains three complex scalar fields $X$ in the
adjoint representation, transforming under the $SO(6)$ R-symmetry, and
we can discuss operators as above made from one of these scalar
fields.
This field theory is dual to type IIB
string theory on $AdS_5\times RP^5$ \witson.

In the simple example above, \free, we saw that if only
$D_{1,1}\neq 0$ our approximation gives the right result, i.e. the
free diagram only. On the other hand, in the $\CN=4$ SYM theory
$\Pf(X)$ is a chiral primary so $\vev{\Pf(X)\Pf(X^\dagger)}$ is
protected. All diagrams cancel out except for the free diagram, so
$a_N=(2D_{1,1})^N$ exactly. This could be taken to suggest that in
this theory all the $D_{k,\chi}$'s vanish except for the
propagator $D_{1,1}$ (which is independent of
the 't Hooft coupling in this case). If this is true then in this 
theory we know
for sure that the saddle point computation is justified, and that
$\beta_0 = 1 / 2D_{1,1}$. However, we have not been able to prove
this.

Assuming that we are studying a supersymmetric theory in which
this is indeed correct,
there is another calculation we can make. The contribution of diagrams
in which the interacting component has $l$ incoming external legs is
proportional to
(taking the propagator to be one for simplicity):
\eqn\susy{
\frac{N!}{N^N}(2N)^{\chi}\frac{N^{N-l}}{(N-l)!},  }
because (with our assumption) the only
contribution comes from having trivial propagators in the rest of the
diagram and these give this symmetry factor.

It is easy to see that diagrams with larger values of $l$ are
suppressed in the large $N$ limit, so that diagrams whose number of
trivial legs is of order $N$ dominate. In fact, from \susy, given the
topology, the dependence of the diagram on $l$ is proportional to
$\frac{N(N-1)...(N-l+1)}{N^{l}}$ (ignoring the $N$-independent factor
$C_{l,\chi}$).  We have
\eqn\distribution{\eqalign{\log(\frac{N(N-1)...(N-l+1)}{N^{l}}) =&
\sum_{j=1}^l \log(1-\frac{j-1}{N}) =
-\frac{\sum_{j=0}^{l-1}j}{N}+O(\frac{l^3}{N^2})= \cr =&
-\frac{l(l-1)}{2N}+O(\frac{l^3}{N^2}),
}}
so the dependence of the $N$-factor of the diagram on $l$ is
proportional to the factor
$Exp(-\frac{l(l-1)}{2N})(1+O(\frac{l^3}{N^2}))$ (when
$l\approx N$ it is easy to see that we get $\approx Exp(-N)$).  This
means that it is sufficient to consider diagrams in which the number of
legs attached to the internal surface is up to a number of order
$\sqrt{N}$, and this is also the expectation value of the number of the
interacting legs.

The internal surface represents the worldsheet in string
theory, and we may think of its boundary as a closed string boundary
state. The internal surface is really part of a discrete Feynman
diagram and not a continuous surface like the worldsheet, so the
boundary state is represented in the CFT as made from a number of
discrete units. There are roughly the same number of vertices on the
boundary as there are external legs, which means that the boundary state
string has $\sim\sqrt{N}$ discrete units.  This is somewhat similar to
\bmn\ where closed string states in the plane wave limit of $AdS_5\times
S^5$ correspond to traces of $\sim\sqrt{N}$ fields in the field
theory, so they are also made of $\sim\sqrt{N}$ discrete units.

\newsec{$SU(N)$ Gauge Theories}

In this section we will deal with $SU(N)$ gauge theories. In these theories,
\ratio\ takes the form
\eqn\predict{\frac{\langle\det_{L}(X)\det_{L}(X^{\dagger})
N\tr(X^{l_{1}}) N\tr(X^{l_{2}}) \cdots N\tr(X^{\dagger m_{1}})
N\tr(X^{\dagger m_{2}}) \cdots
\rangle}{\langle\det_{L}(X)\det_{L}(X^{\dagger})\rangle}.}
As discussed in section 3, we expect \predict\ to have a topological
large $N$ expansion which will include all oriented surfaces with
boundaries (at least for the $\CN=4$ SYM theory, which is equivalent
to a string theory), and we will develop this expansion in this
section. The analysis will basically parallel that of the $SO(2N)$ case
(section 4) with a few changes.

The outline of this section is as follows: in \S5.1 we will review the
usual  large $N$  expansion  of  this theory,  which  applies only  to
correlators  of traces of  small powers  of $X$.   In \S5.2  we will
analyze   the  form  and   $N$-dependence  of   a  general   diagram  in
$\langle\det_L(X)\det_L(X^{\dagger})\rangle$, and  in \S5.3 we  will extend the
analysis  to  diagrams  for   correlators  which  also  include  trace
operators.  We  will  identify  the topology  corresponding  to  every
diagram and find that all oriented topologies, including those with boundaries,
appear.

The analysis will make use only of the general color structure
of the theory. The main difference from the $SO(2N)$ case is in the
different constraints on the indices in the definitions of
$\det_L(X)$. One of the results will be the appearance of a class of
diagrams which do not exist for $SO(2N)$, and it will be natural to
interpret them as processes with intermediate brane states of angular
momentum different from $L$, such as $det_M(X), \ M \neq L$ (similar
subleading diagrams do not exist for Pfaffians in the $SO(2N)$ theory
and this is consistent with the fact that there are no Pfaffian-like
states with angular momentum less than $N$).
Note that we could also analyze
operators of the type $\det_L(X)$ in the $SO(2N)$ gauge theory, and we
would find similar results to the ones we present here.

In \S5.4 we will discuss the large $N$ expansion of \predict. We will
argue that the contribution of all diagrams with a specific topology
is proportional to $N^{\chi}$, where $\chi$ is the Euler characteristic of
that topology. This will justify our identification of the topology of
each Feynman diagram. The discussion of factorization here is completely
identical to the $SO(2N)$ case, so we will not repeat it.

\subsec{Review and definitions}

Let $X$ be a scalar field in the adjoint representation of $SU(N)$
(the $\CN=4$ SYM theory contains three such complex R-charged fields).  We may
label $X$ by one fundamental and one anti-fundamental index,
$X_{i}^{j}$, where $i,j=1,\cdots,N$. In this notation, the operators of
interest to us are the traces
\eqn\Trop{\tr(X^L)=X_{i_2}^{i_1}X_{i_3}^{i_2} X_{i_4}^{i_3}\cdots
X_{i_1}^{i_L}} with $L \ll N$, and the subdeterminant operators
\eqn\Sdetopp{\det_L(X)=\frac1{L!(N-L)!}  \epsilon_{i_1\cdots
i_Li_{L+1}\cdots i_N} \epsilon^{j_1\cdots j_Li_{L+1}\cdots i_{N}}
X_{j_{1}}^{i_{1}}\cdots X_{j_{L}}^{i_{L}} }
with $N-L \ll N$.
On the string theory side, the operators \Trop\ correspond to closed string
states, gravitons and their superpartners, with small angular momentum
$L$ ($\ll N$) on $S^{5}$, and the operators \Sdetopp\ correspond to
``giant gravitons'', states with large angular momentum $L$ ($N-L\ll
N$) on $S^{5}$, which become spherical D3-branes. As explained in
section 3, based on string perturbation theory we expect correlators
of closed string states to have a large $N$ topological expansion with
only even powers of $1/N$ (corresponding to oriented closed surfaces),
while correlators of closed string states with ``giant gravitons''
should have a large $N$ topological expansion with even as well as odd
powers of $1/N$ (corresponding to oriented surfaces with boundaries).

The 't Hooft large $N$ expansion (in which $\lambda=Ng_{YM}^2$ is held
fixed) of correlators of usual single-trace operators is well known
\thooft\ and we will briefly review the essential points.

\fig{The double line notation: (a) the propagator (b) a vertex (c) the
free planar diagram for $\langle\tr (X^3) \tr (X^{{\dag}3})\rangle$.
}{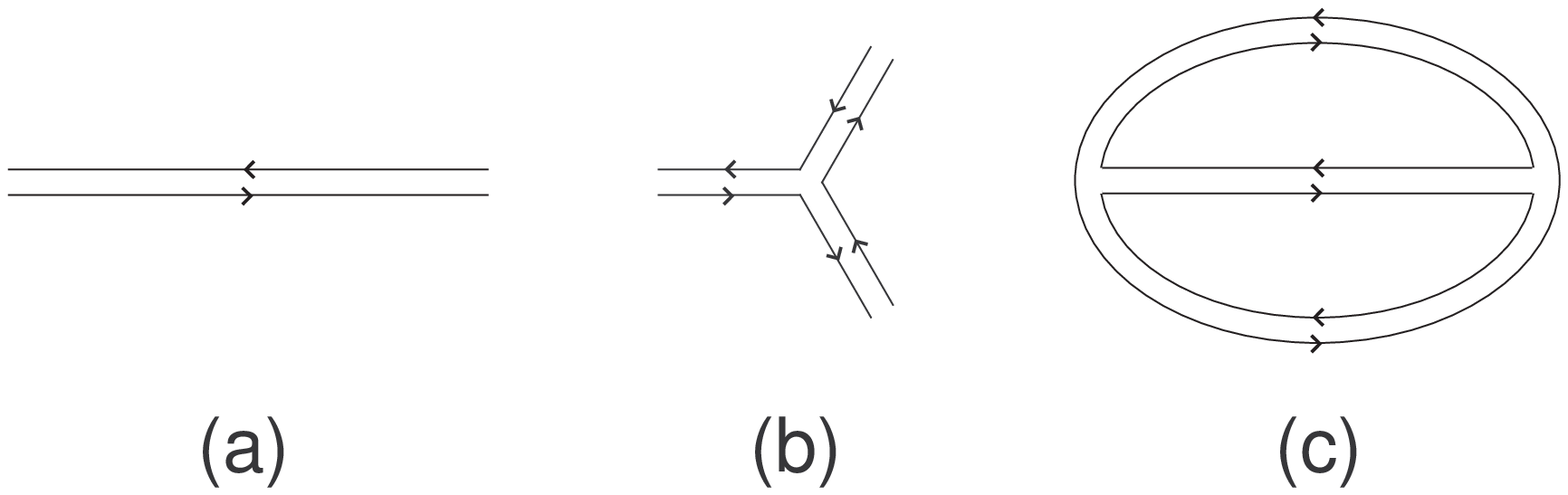}{4truein}
\figlabel{\doubleline}

The Feynman diagrams of an $SU(N)$ gauge theory with adjoint fields
can be written in the double line notation, in which the propagator
$\langle X^{i}_{j}X^{\dagger k}_{l}\rangle$ is represented by two
directed lines, one with the fundamental index and one with the
anti-fundamental index (figure \doubleline(a)). The propagator is given by
\eqn\propSUN{\langle X^{i}_{j}(x_1) X^{\dagger k}_{l}(x_2) \rangle
\propto {{\delta^{i}_{l}\delta^{k}_{j}-
\frac{1}{N}\delta^{i}_{j}\delta^{k}_{l}}\over {(x_1-x_2)^2}},} so to
leading order in $1/N$, the incoming and outgoing indices on each line
in a propagator are equal.
The interaction terms that appear in the Lagrangian are single traces
(see figure \doubleline(b)).
When a closed loop is formed in the diagram the index
runs over all possible values and contributes a factor of $N$ to the
value of the diagram.
It is
convenient to normalize the fields such that there is a factor of
$1/g_{YM}^2 =
N/\lambda$ in front of the whole Lagrangian and no other dependence
on $g_{YM}$. Then, each vertex in a Feynman diagram gives a factor of
$N/\lambda$ and each propagator gives a factor of $\lambda/N$.

Thus, the $N$ dependence of a vacuum Feynman diagram in the large $N$
limit with fixed $\lambda$ is, to leading order in $1/N$, $N^{V-E+F}$,
where $V$ is the number of vertices, $E$ the number of propagators,
and $F$ the number of closed loops in the diagram. We may think of the
diagram as a triangulation of a two-dimensional surface, with a face
corresponding to each closed loop, and an edge to each (double lined)
propagator. The leading $N$ power of the diagram, $V-E+F$, is then
equal to the Euler characteristic of the surface. A topological
theorem states that every compact two-dimensional oriented surface is
homeomorphic to a sphere with a certain number of handles and
boundaries (holes), and that all triangulations of the surface have
the same value of $V-E+F$, given by $\chi=2-2H-B$ where $H$ is the
number of handles and $B$ is the number of boundaries.

Incoming and outgoing graviton states correspond to trace operators
similar to those appearing as interaction vertices.  Hence, the power
of $1/N$ corresponding to a diagram involving such states can be
calculated in the same way as for vacuum diagrams, replacing the
graviton insertions by vertices. All the surfaces constructed in this way
are closed, i.e. without boundaries, so their Euler characteristics are
even. Thus, the large $N$ expansion only includes even powers of
$N$. The leading diagrams will correspond to spheres (which means they
are planar), followed by the torus and so on.

\subsec{The $\langle\det_{L}(X)\det_{L}(X^{\dagger})\rangle$ correlation 
function}

\fig{The general form of a Feynman diagram for
$\langle\det_{L}(X)\det_{L}(X^{\dagger})\rangle $.
}{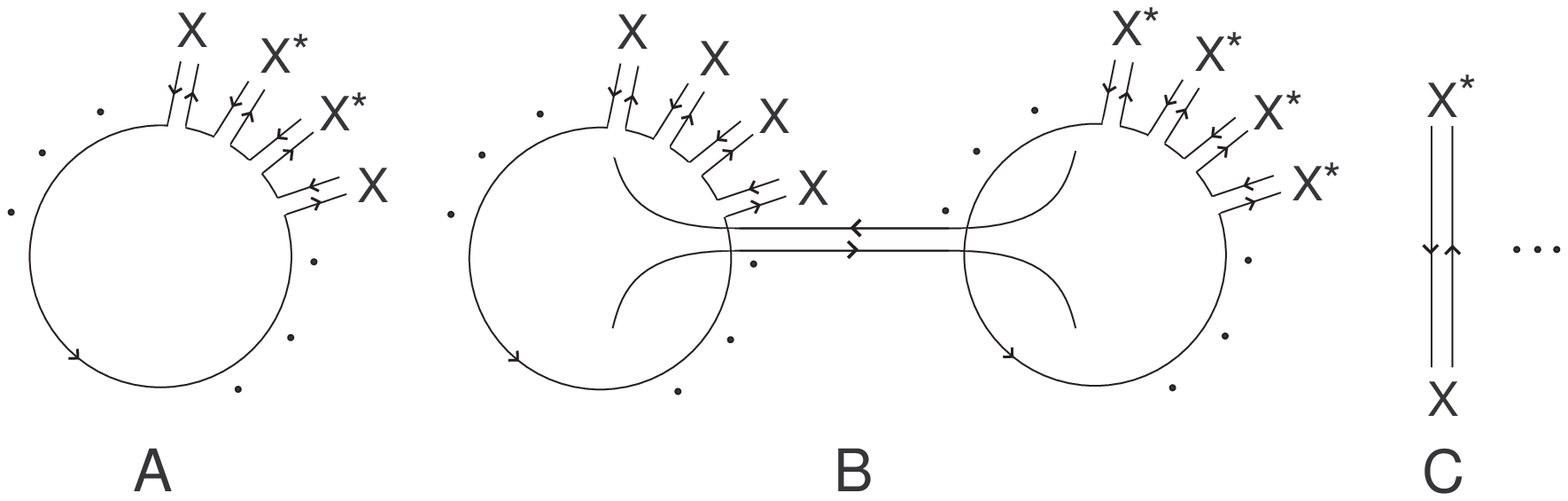}{5truein}
\figlabel{\generaldet}

We will begin with $\langle\det_{L}(X)\det_{L}(X^{\dagger})\rangle $
with no traces in the correlator. The general form of a Feynman
diagram for this correlator is shown in figure \generaldet. Using similar terms
to the $SO(2N)$ case, it consists of several connected components,
each having a number of external legs connected to it, but in this
theory they are attached in an arbitrary order. Unlike the $SO(2N)$
case, the external legs do not have to alternate between the incoming
and outgoing determinants.

\fig{Amputating the diagrams. Boundaries are denoted by dashed
lines. (a) A disk diagram. (b) A torus with one hole. (c) A cylinder
(two holes in one connected component).  }{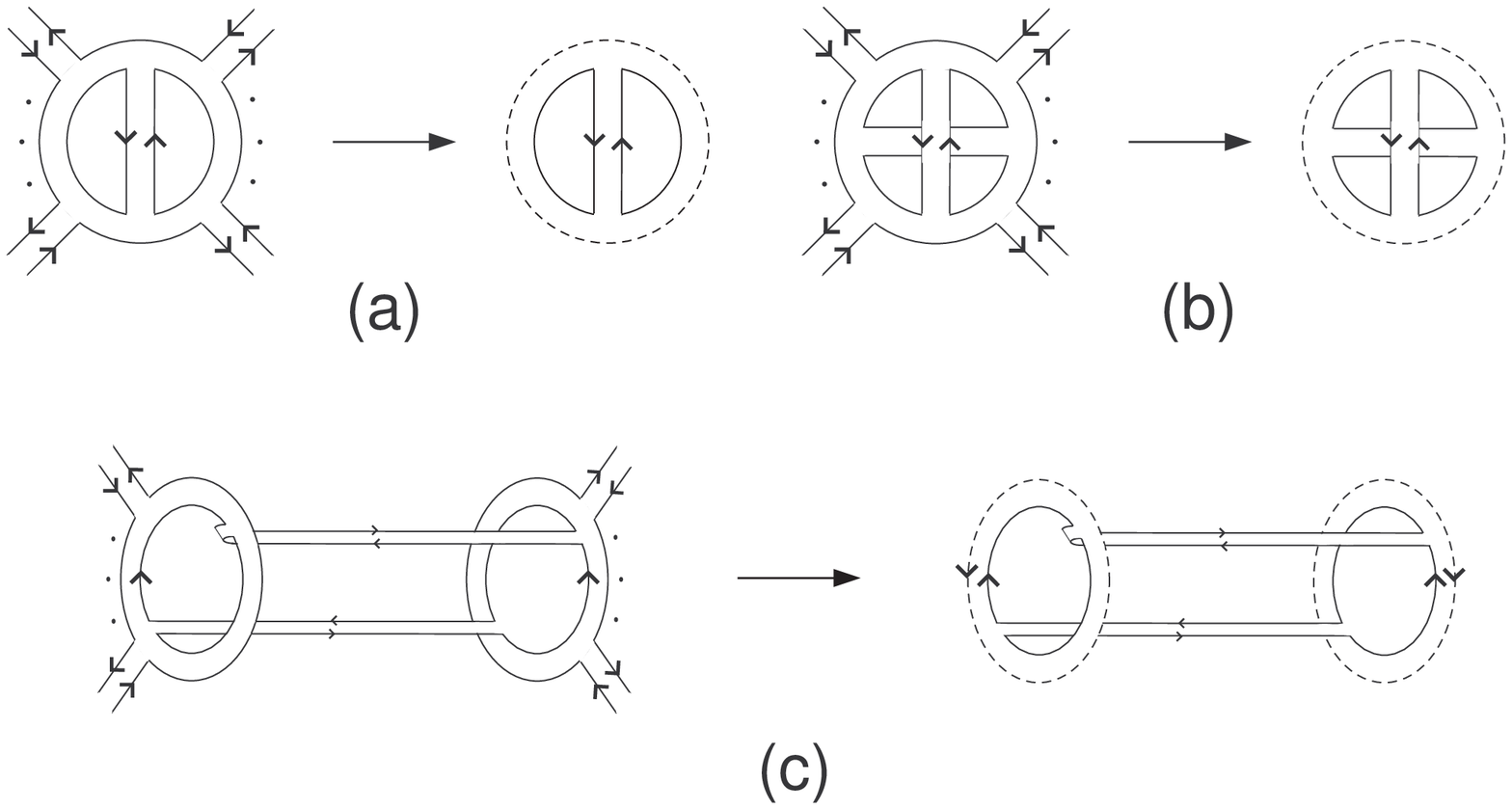}{5truein}
\figlabel{\amputationsu}

We begin by developing some terminology. First, just as in the $SO(2N)$
theory, imagine that we amputate the lines coming from external legs.
The remaining edges and vertices form a
vacuum diagram which contains only trace vertices (figure \amputationsu). The
diagram forms a surface which can be treated using the usual large $N$
expansion. As in the $SO(2N)$ case, we will
regard each amputated line
as (part of) a boundary for this surface, and we will call these surfaces the
internal surfaces of the diagram.  As in section 4, we will consider a
trivial component (one propagator running from $X$ to $X^{\dagger}$)
to have the internal topology of a disk.

\fig{Types of external lines. The lines can be divided into
four types according to their origin and destination.
}{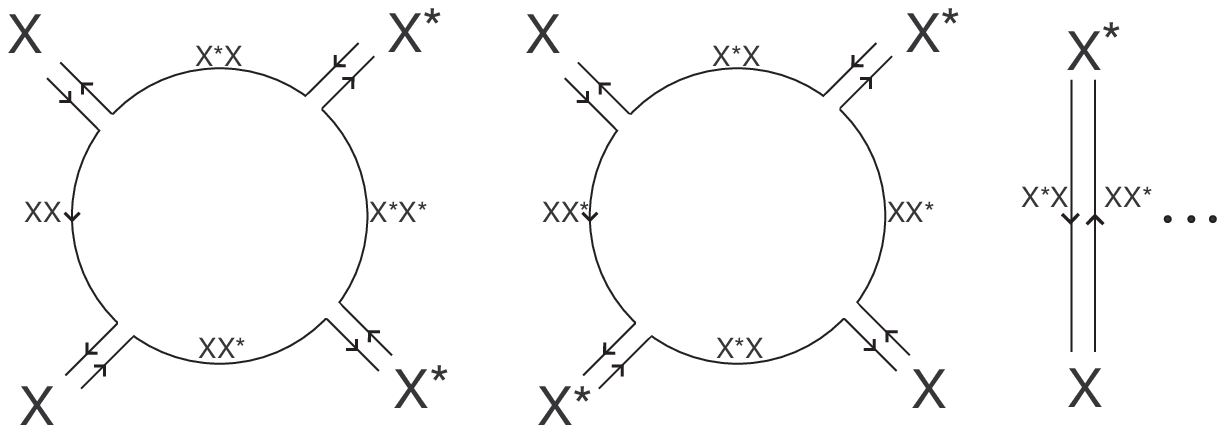}{5truein}
\figlabel{\indextypes}

Second, we have already noted that in this theory, the external legs
are attached to the internal surfaces in an arbitrary order, unlike
the $SO(2N)$ case where there was only one way to connect them because of
the antisymmetric nature of the Pfaffian operators. In the double line
notation, each
propagator is represented by two lines directed in opposite directions
and carrying one index each.  Since there are $L$ incoming external
legs and $L$ outgoing external legs in a diagram, there are $2L$
directed external lines in the diagram. Those lines can be of one of
four types: they either go from $X$ to $X^{\dagger}$, from $X$ to $X$,
from $X^{\dagger}$ to $X$ or from $X^{\dagger}$ to $X^{\dagger}$. Let
$L_{XX^{\dagger}}$, $L_{XX}$, $L_{X^{\dagger}X}$, and
$L_{X^{\dagger}X^{\dagger}}$ denote the number of lines (and hence
indices) belonging to each type, respectively. It is clear (see figure
\indextypes) that $L_{XX}=L_{X^{\dagger}X^{\dagger}}$ and
$L_{XX^{\dagger}}=L_{X^{\dagger}X}$.

What are the possible values of $L_{XX^{\dagger}}$ ?  Each trivial
component in the diagram contributes one to $L_{XX^{\dagger}}$.
The other external propagators connect to the
boundaries of the internal surfaces. There are several ways to attach
them, each giving a different value of $L_{XX^{\dagger}}$.  When the
external legs are arranged in alternating order (one from $\det_{L}(X)$,
one from $\det_{L}(X^{\dagger})$, and so on) $L_{XX^{\dagger}}$ will be
equal to its maximal value, $L$. Each swap of neighboring external
legs reduces $L_{XX^{\dagger}}$ by one.

Consider a diagram with $L_{XX^{\dagger}}=L$. In such a diagram, each
cut in the diagram (between the $\det_L(X)$ and $\det_L(X^{\dagger})$ vertices)
will cross at least $L$ pairs of (single index) external
lines. In a diagram with $L_{XX^{\dagger}}=L-1$, however, it is
possible to cut the diagram in such a way that the cut will cross only
$L-1$ pairs of external lines.  When $L_{XX^{\dagger}}$ decreases
further, there are cuts which cross smaller numbers of line pairs. It
is natural to interpret diagrams which have cuts crossing $K$ pairs of
external lines as diagrams which have intermediate states involving
$\det_K(X)$, or in string theory language, processes which have ``giant
gravitons'' with a lower angular momentum $K$ as intermediate
states. The fact that similar diagrams do not exist for Pfaffians in the
$SO(2N)$ theory is consistent with the fact that there are no
Pfaffian-like states with angular momentum less than $N$.

In any case, we will see that the above characteristics of a general
diagram, the topology of the internal surfaces and the value of
$L_{XX^{\dagger}}$, fully determine its $N$-dependence.

\fig{The free diagram for
$\langle\det_{L}(X)\det_{L}(X^{\dagger})\rangle $. The difference from
the $SO(2N)$ case is that the propagators have direction.
}{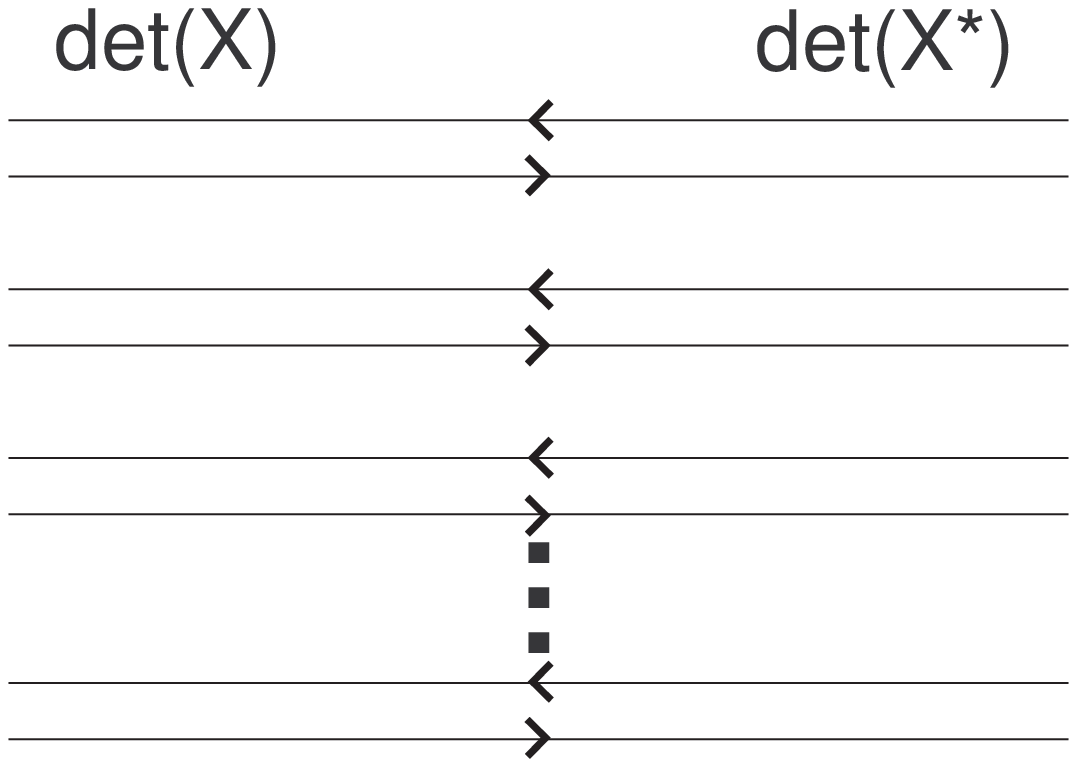}{2.3truein}
\figlabel{\freedet}

As a simple example, we start with the calculation of the $N$
dependence of the free diagram shown in figure \freedet. First we
note that the $\frac{1}{L!}$ factor in the definition of
$\det_{L}(X)$ can be cancelled by all the different permutations
of the incoming external legs. The same is true for
$\det_{L}(X^\dagger)$ and permutations of outgoing external legs.
However, when we permute the legs of both $X$ and $X^\dagger$
there are $L!$ permutations which give the same diagram. Since we
want to count each diagram only once we need to divide by this
symmetry factor.

Every line has an index which is summed over, but the indices are
constrained by the definition \Sdetop\ in a complicated way. We
therefore need to count the different ways to assign indices to
every line. Since the lines are directed we can divide them into
two types. Lines starting on the incoming subdeterminant and
ending on the outgoing subdeterminant we call the $XX^\dagger$
type, and lines starting on the outgoing subdeterminant and ending
on the incoming subdeterminant we call the $X^\dagger X$ type (as
shown in figure \indextypes). The definition of the subdeterminant
implies that the group of the $L$ indices of type $XX^\dagger$ and
the group of the $L$ indices of type $X^\dagger X$ must be the
same set of different numbers. To choose such a set there are
$\frac{N!}{(N-L)!L!}$ ways, and there are $(L!)^2$ ways of
assigning them to the different lines. The $L$ propagators give
$N^{-L}$.
Putting it all together we get
\eqn\freesun{
\frac1{L!} \frac{N!}{(N-L)!L!} L!L! N^{-L} ={{N!} \over {(N-L)!}} N^{-L}.
}

We now proceed to the calculation of a general diagram. As in the
$SO(2N)$ theory, there are three sources for the $N$ dependence of a
diagram. The analysis of the first two parts is
identical to the $SO(2N)$ theory, but the third part will be
different.

\item{1.} External propagators: there are $2L$ external propagators
and they give a factor of $N^{-2L}$. Also, the $\frac{1}{L!}$ factors
in the definitions of $\det_{L}(X)$ cancel the various permutations of
the external legs, up to whatever symmetry factor the diagram has:
every group of $C$ identical components in a diagram gives a
$\frac1{C!}$ symmetry factor.

\item{2.} Internal surfaces: the edges, vertices and loop indices of
the internal surfaces contribute $N^\chi$ each, where $\chi$ is the
Euler characteristic of the internal topology (recall that we consider each
trivial component to be a disk). The definition of the boundaries of
the surfaces means that we have not taken into account the counting of
the indices which run on these boundaries. We will do so now.

\item{3.} Counting the external indices: the count depends on how the
external legs connect to the boundaries of the internal surfaces.
The definition \Sdetopp\ implies that all incoming lines must carry
different indices, and the same for the outgoing lines, and that
these two index sets must be the same.
For the general diagram there are $\frac{N!}{(N-L_{XX^{\dagger}})!}$
ways to choose the $XX^{\dagger}$ indices. The $XX$ indices must all
be different from these, so we can choose them in
$\frac{(N-L_{XX^{\dagger}})!}{(N-L)!}$ different ways. The same
applies to the $X^{\dagger}X^{\dagger}$ indices, which are independent
of the $XX$ indices, and there are
$\frac{(N-L_{XX^{\dagger}})!}{(N-L)!}$ ways to choose them as well. We
still have to choose the $X^{\dagger}X$ indices.  Since the incoming
and outgoing indices on each side must be the same set, and the
$X^{\dagger}X$ indices must be different from both the $XX$ indices
and the $X^{\dagger}X^{\dagger}$ indices, it follows that the set of
$X^{\dagger}X$ indices must equal the set of $XX^{\dagger}$
indices. This leaves us with just $L_{XX^{\dagger}}!$ possible
permutations. Altogether we get
\eqn\count{\frac{N!}{(N-L_{XX^{\dagger}})!}
\left(\frac{(N-L_{XX^{\dagger}})!}{(N-L)!}\right)^{2}L_{XX^{\dagger}}!
=\frac {N! (N-L_{XX^{\dagger}})! L_{XX^{\dagger}}!} {(N-L)!^{2}}.}

Combining the three parts together, we get
\eqn\withf{\frac1{\rm{Symmetry\ Factor}} N^{-2L} N^{\chi}
\frac{N!(N-L_{XX^{\dagger}})!L_{XX^{\dagger}}!}{(N-L)!^{2}}.}

It will be convenient to normalize by dividing by the free diagram
\freesun. This gives
\eqn\withfnorm{\frac{N^{-L}}{(N-L)!} \frac1{{\rm{Symmetry\ Factors}}}
(N-L_{XX^{\dagger}})! L_{XX^{\dagger}} !N^{\sum\chi},} where
$\sum\chi$ is the sum of the Euler characteristics of the internal
surfaces.

\subsec{Correlation functions with traces}

\fig{Some possible diagrams appearing in the computation of the
correlation function
$\langle\det_L(X)\det_L(X^{\dagger}) N\tr (X^l) N\tr (X^{\dagger
l})\rangle $. 
}{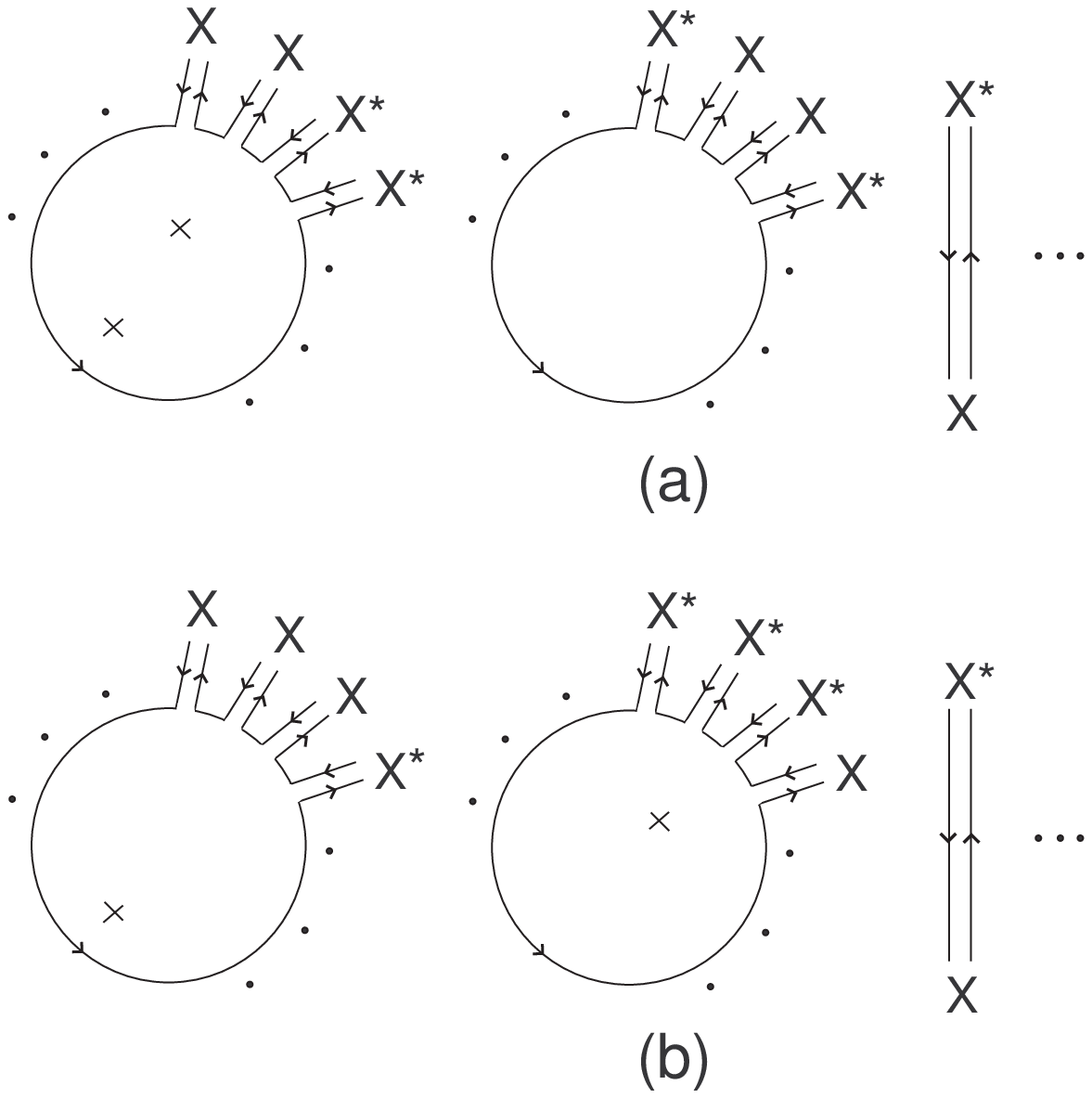}{4truein}
\figlabel{\generaldettr}

Consider the correlator
\eqn\constLcorr{
\langle\det_{L}(X)\det_{L}(X^{\dagger})
N\tr(X^{l_{1}}) N\tr(X^{l_{2}}) \cdots N\tr(X^{\dagger m_{1}})
N\tr(X^{\dagger m_{2}}) \cdots \rangle.} 
We can think of the single-trace 
operators as internal interaction vertices appearing in the
diagram rather than external incoming or outgoing states. The general
diagram for this correlator will therefore look something like those in figure
\generaldettr. Just like the diagrams in
$\langle\det_{L}(X)\det_{L}(X^{\dagger})\rangle $, it consists of
several components, some of which contain trace insertions.  The $N$
dependence of these diagrams is identical to that of the same diagram
in $\langle\det_L (X) \det_L(X^{\dagger})\rangle $, but with the trace
insertions replaced by ordinary interaction vertices, so it is given
(normalized by the free diagram) by \withfnorm.

\subsec{Expansion in topologies}

In this section we will develop the large $N$ expansion for \predict.
The ideas are identical to those of \S4.4. We identify the internal
topology of the interacting component (the set of components containing
the trace insertions) of each diagram with the corresponding
topology in string theory. To justify this, we will show that the leading
contribution of all diagrams with topologies of Euler characteristic
$\chi$ is proportional to $N^\chi$.  We will begin by writing expressions
for the denominator and the numerator of \predict, and then use the
saddle point method to obtain the desired result for their ratio.
The calculation will be very similar to the one in \S4.4, but somewhat
more complicated, as there are more classes of diagrams to consider.
Here too, we will need to make some assumptions on the existence of
certain integration contours in order to use the saddle-point method.

\medskip
\noindent {{\it 5.4.1} \bf $\langle\det_L(X)\det_L(X^{\dagger})\rangle$}
\medskip

Each connected component in a $\langle\det_L(X)\det_L(X^{\dagger})\rangle$
diagram can be characterized by $k_1$, the number of incoming
external legs attached to it, $k_2$, the number of outgoing legs,
$s$, the number of lines which go from $X$ to $X^\dagger$ (recall the
discussion of $L_{XX^{\dagger}}$ in \S5.2), and $\chi$, the Euler
characteristic of the internal topology (in the $SO(2N)$ case, the fact
that the legs attached to each component must be arranged in an alternating
order implied that $k_1=k_2$, but in the $SU(N)$ theories this
need not be the case). Let $D_{k_1,k_2,\chi,s}$ denote the sum of all
connected diagrams which form a component of type $(k_1,k_2,\chi,s)$, but
without the $N$-dependent factors.

Consider first the sum of all diagrams which have $L_{k_1,k_2,\chi,s}$
connected components of type $(k_1,k_2,\chi,s)$. Normalized by the free
diagram it is given by:
\eqn\sunonediagram{\frac1{(N-L)!N^{L}}(N-S)!S!
\prod_{k_1,k_2,\chi,s}
\frac{N^{\chi L_{k_1,k_2,\chi,s}}D_{k_1,k_2,\chi,s}^{L_{k_1,k_2,\chi,s}}}
{L_{k_1,k_2,\chi,s}!},}
where $S=\sum_{k_1,k_2,\chi,s} s L_{k_1,k_2,\chi,s}$ is
$L_{XX^{\dagger}}$ in the notation of \S5.2 (recall \withfnorm). The
product extends over all $k_1,k_2=0,\cdots,N; \chi=1,0,-1,\cdots;
s=0,\cdots,{\rm min}(k_1,k_2)$. 
It is easy to see that the factorial terms in the
denominators give the right symmetry factors for each diagram in the sum.

To compute $\vev{\det_L(X) \det_L(X^{\dagger})}$ we sum 
over all possibilities for $L_{k_1,k_2,\chi,s}$, obtaining
\eqn\sunalldiagramsnoUsym{\sum_{L_{k_1,k_2,\chi,s}\geq 0}
\frac1{(N-L)!N^{L}}(N-S)!S!\prod_{k_1,k_2,\chi,s}\frac{N^{\chi
L_{k_1,k_2,\chi,s}}D_{k_1,k_2,\chi,s}^{L_{k_1,k_2,\chi,s}}}
{L_{k_1,k_2,\chi,s}!},}
where the sum is constrained by the two conditions
$\sum_{k_1,k_2,\chi,s}k_1 L_{k_1,k_2,\chi,s}=L$ and
$\sum_{k_1,k_2,\chi,s}k_2 L_{k_1,k_2,\chi,s}=L$.
We will denote the same sum \sunalldiagramsnoUsym\ with the constraints
$\sum_{k_1,k_2,\chi,s}k_1 L_{k_1,k_2,\chi,s}=n$ and
$\sum_{k_1,k_2,\chi,s}k_2 L_{k_1,k_2,\chi,s}=m$ by $a_{nm}$.

\medskip
\noindent {{\it 5.4.2} \bf $\langle\det_L(X)\det_L(X^{\dagger})
N\tr(X^{j_1})N\tr(X^{j_2})\cdots
N\tr({X^{\dagger}}^{m_1})N\tr({X^{\dagger}}^{m_2})\cdots\rangle$}
\medskip

Turning our attention to correlation functions with insertions,
let us evaluate the contribution of all diagrams whose interacting
component has Euler characteristic $\chi_0$ \foot{As in the
$SO(N)$ theory, the interacting `component' may be composed of several
disconnected components.}.
Every such diagram is characterized by the values of $l_1,l_2$ (the
number of incoming and outgoing legs of the interacting component),
$s_0$ (the contribution of its interacting component to $S$), and as
before, by $L_{k_1,k_2,\chi,s}$, the number of (insertion free)
components of type $(k_1,k_2,\chi, s)$. These satisfy the two conditions
$\sum_{k_1,k_2,\chi,s}k_1 L_{k_1,k_2,\chi,s}+l_1=L$ and
$\sum_{k_1,k_2,\chi,s}k_2 L_{k_1,k_2,\chi,s}+l_2=L$.

Let $C_{l_1,l_2,\chi,s}$ denote the sum of all diagrams of type
$(l_1,l_2, \chi, s)$ which include the trace
insertions but with no $N$ dependence.  The contribution of all diagrams
that have specific values of $l_1,l_2$, $s_0$ and $L_{k_1,k_2,\chi,s}$ is
\eqn\sunonediagramtrace{\frac1{(N-L)!N^{L}} N^{\chi_0} C_{l_1,l_2,\chi_0,s_0}
(N-S-s_0)!(S+s_0)!\prod_{k_1,k_2,\chi,s}\frac{N^{\chi
L_{k_1,k_2,\chi,s}}D_{k_1,k_2,\chi,s}^{L_{k_1,k_2,\chi,s}}}
{L_{k_1,k_2,\chi,s}!}.}

The total contribution comes from summing over all values of $l_1,l_2$,
$s_0$ and $L_{k_1,k_2,\chi,s}$:
\eqn\sunalldiagramstracenoUsym{\sum_{L_{k_1,k_2,\chi,s};l_1,l_2,s_0}
\frac{N^{\chi_0} C_{l_1,l_2,\chi_0,s_0}}{(N-L)!N^{L}}
(N-S-s_0)!(S+s_0)!\prod_{k_1,k_2,\chi,s}
\frac{N^{\chi
L_{k_1,k_2,\chi,s}}D_{k_1,k_2,\chi,s}^{L_{k_1,k_2,\chi,s}}}
{L_{k_1,k_2,\chi,s}!},}
where the sum is constrained by the two constraints
$\sum_{k_1,k_2,\chi,s}k_1 L_{k_1,k_2,\chi,s}+l_1=L$ and
$\sum_{k_1,k_2,\chi,s}k_2 L_{k_1,k_2,\chi,s}+l_2=L$.
We will denote the same sum \sunalldiagramstracenoUsym\ with the constraints
$\sum_{k_1,k_2,\chi,s}k_1 L_{k_1,k_2,\chi,s}+l_1=n$ and
$\sum_{k_1,k_2,\chi,s}k_2 L_{k_1,k_2,\chi,s}+l_2=m$ by $b_{nm}$.

\medskip
\noindent {\it 5.4.3 Saddle point evaluation}
\medskip

We want to show that the normalized correlation function, the
ratio $\frac{\sunalldiagramstracenoUsym}{\sunalldiagramsnoUsym}=\frac{b_{LL}}{a_{LL}}$,
is proportional to $N^{\chi_0}$. We would like to evaluate it
using the saddle point method, as in \S4.4. One complication in
this theory is that the appearance of the $S!(N-S)!$ factors makes
it more difficult to get a useful form for the partition
functions. Using the Beta function, however, we will be able to
write the partition function as an integral of a simple
exponential function of the kind we obtained in \S4.4, at the price of
needing to perform an additional saddle point integration.

Another complication arises from the fact that the number of
incoming and outgoing legs of a component need not be equal. Thus, we
have two conditions in each sum (\sunalldiagramsnoUsym\ and
\sunalldiagramstracenoUsym), so the
partition function will have two arguments and a double
integration will be necessary. We therefore define the `partition
functions' :
\eqn\sunpartition{Z_1(\beta,\gamma)\equiv
 \sum_{n,m=0}^{\infty} a_{nm} \beta^n \gamma^m, \quad
Z_2(\beta,\gamma)\equiv
 \sum_{n,m=0}^{\infty} b_{nm} \beta^n \gamma^m
.}

From the definition of the Beta function it follows that
\eqn\betafunction{
S!(N-S)!=(N+1)!B(S+1,N-S+1)=(N+1)!\int_0^1 dx x^S (1-x)^{N-S}.}
Inserting this into \sunalldiagramsnoUsym,
\sunalldiagramstracenoUsym\ and \sunpartition\ we get
\eqn\sunalldiagramsreg{\eqalign{ Z_1(\beta,\gamma) = &
\sum_{L_{k_1,k_2,\chi,s}\geq 0}\frac1{(N-L)!N^{L}}(N-\sum_{k_1,k_2,\chi,s}s
L_{k_1,k_2,\chi,s})!(\sum_{k_1,k_2,\chi,s}s L_{k_1,k_2,\chi,s})!
\cdot
\cr &\cdot \prod_{k_1,k_2,\chi,s}\frac{N^{\chi
L_{k_1,k_2,\chi,s}} 
\beta^{k_1 L_{k_1,k_2,\chi,s}} \gamma^{k_2 L_{k_1,k_2,\chi,s}}
D_{k_1,k_2,\chi,s}^{L_{k_1,k_2,\chi,s}}}{L_{k_1,k_2,\chi,s}!}
\cr = & \int_0^1 dx
\sum_{L_{k_1,k_2,\chi,s}\geq 0}\frac{(N+1)!}{(N-L)!N^{L}} x^{N}
{(\frac{1-x}{x})}^{(\sum_{k_1,k_2,\chi,s}s L_{k_1,k_2,\chi,s})}
\cdot
\cr & \cdot \prod_{k_1,k_2,\chi,s}\frac{N^{\chi
L_{k_1,k_2,\chi,s}}
\beta^{k_1 L_{k_1,k_2,\chi,s}} \gamma^{k_2 L_{k_1,k_2,\chi,s}}
D_{k_1,k_2,\chi,s}^{L_{k_1,k_2,\chi,s}}}{L_{k_1,k_2,\chi,s}!} \cr = &
\frac{(N+1)!}{(N-L)!N^{L}}
\int_0^1 dx\ x^{N} e^{\sum_{k_1,k_2,\chi,s}N^{\chi} D_{k_1,k_2,\chi,s}
\beta^{k_1}\gamma^{k_2} (\frac{1-x}{x})^{s}},
}}
and similarly
\eqn\sunpartitiontrace{Z_2(\beta,\gamma) =
\frac{(N+1)!}{(N-L)!N^{L}}N^{\chi_0}
\int_0^1 dx\ x^{N} e^{\sum_{k_1,k_2,\chi,s}N^{\chi} D_{k_1,k_2,\chi,s}
\beta^{k_1}\gamma^{k_2} (\frac{1-x}{x})^{s}}
g_{\chi_0}(\beta,\gamma,x),
}
where
\eqn\g{g_{\chi_0}(\beta,\gamma,x)\equiv
\sum_{l_1,l_2,s_0} C_{l_1,l_2,\chi_0,s_0} \beta^{l_1}\gamma^{l_2}
\left( \frac{1-x}{x} \right)^{s_0}.}

The coefficients $a_{LL}$ and $b_{LL}$
are given by the double contour integrals:
\eqn\sunintegralN{\eqalign{
a_{LL}=-\frac{1}{4\pi^2} \oint d\beta \oint d\gamma \ Z_1(\beta,\gamma)
\beta^{-(L+1)}\gamma^{-(L+1)} ,
\cr
b_{LL}=-\frac{1}{4\pi^2} \oint d\beta \oint d\gamma \ Z_2(\beta,\gamma)
\beta^{-(L+1)}\gamma^{-(L+1)} .}}

We will begin by analyzing $a_{LL}$.  Recall that we are interested in
$L$'s such that $(N-L)$ remains fixed as $N\rightarrow
\infty$, so we will treat $(N-L)$ as an object of order $1$.
Keeping only the two leading (first and zeroth) powers of $N$
in the exponent of $Z_1(\beta,\gamma)$ (the leading term cannot vanish since
the theory has a propagator) we get
\eqn\sunintegralNapprox{a_{LL}\simeq
-\frac{1}{4\pi^2}
\frac{(N+1)!}{(N-L)!N^{L}}
\oint \frac{d\beta}{\beta^{(L-N+1)}}\oint \frac{d\gamma}{\gamma^{(L-N+1)}}
\int_0^1 dx e^{f_0(\beta,\gamma,x)} x^{N} e^{Nf(\beta,\gamma,x)},}
where we define $f(\beta,\gamma,x)\equiv\sum_{k_1,k_2,s}\beta^{k_1}
\gamma^{k_2} (\frac{1-x}{x})^{s}
D_{k_1,k_2,1,s} - \log(\beta)-\log(\gamma)$ to include the leading terms
and $f_0(\beta,\gamma,x)\equiv\sum_{k_1,k_2,s}\beta^{k_1} \gamma^{k_2}
(\frac{1-x}{x})^{s} D_{k_1,k_2,0,s}$ to include the subleading terms;
both functions are independent of $N$.

We will assume that we may exchange the order of integration, and
for each $x$ we perform a saddle point integration on $\beta$ and
$\gamma$. Starting with the $\beta$ integration, we are making the
assumption, as in \S4.4,  that we can always (at least for generic
values of $x$ and $\gamma$) find a contour of fixed imaginary value encircling
the origin which goes through a leading saddle point, say
$\beta_0(\gamma,x)$. The $\beta$ integration then gives
\eqn\betaintegration{
a_{LL}\simeq -\frac{1}{4\pi^2}\frac{(N+1)!}{(N-L)!N^{L}} \int_0^1 dx x^{N}
\oint \frac{d\gamma}{\gamma^{(L-N+1)}}
\frac{e^{i\alpha(\gamma,x)} \sqrt{2\pi}e^{N|f(\beta_0(\gamma,x),\gamma,x)|}
e^{f_0(\beta_0(\gamma,x),\gamma,x)}}
{\sqrt{2N|\partial_\beta^2f(\beta_0(\gamma,x),\gamma,x)|}
\beta_0(\gamma,x)^{L-N+1}}, }
where $e^{i\alpha}$ is a phase determined by the direction of steepest
descent. Similarly, assuming that we can find a steepest descent contour 
around the origin for the $\gamma$ integration we find
\eqn\gammaintegration{\eqalign{
a_{LL}\simeq& -\frac{1}{4\pi^2}\frac{(N+1)!}{(N-L)!N^{L}} \int_0^1 dx x^{N}
e^{i\tilde\alpha(x)} \frac1{\beta_0(\gamma_0(x),x)^{L-N+1}
\gamma_0(x)^{L-N+1}}
\cr & \cdot \frac{2\pi e^{N|f(\beta_0(\gamma_0(x),x),\gamma_0(x),x)|}
e^{f_0(\beta_0(\gamma_0(x),x),\gamma_0(x),x)}}
{2N\sqrt{|\partial_\beta^2f(\beta_0(\gamma_0(x),x),\gamma_0(x),x)
\partial_\gamma^2f(\beta_0(\gamma_0(x),x),\gamma_0(x),x)|}}.
}}

The phase $e^{i\tilde\alpha}$
in the last expression actually makes it real. Indeed,
working in the Euclidean theory, all $D_{k_1,k_2,\chi,s}$ can be
taken to be real, and for each fixed $x$, the $\beta$ and $\gamma$
integrations
just give a coefficient in the power series of a real function.
The important thing about the $x$ integral in \gammaintegration\ is
that it can be written in the form $ \int_0^1 dx u(x) e^{Nv(x)}$ where $u$ and
$v$ are real. This means, since $N\gg 1$, that practically the only
contribution will come from a very narrow part of the unit
interval surrounding the point for which $v(x)$ is maximal (or
possibly diverging), say $x_0$. The exact calculation depends on
whether $x_0$ is inside the interval or on the endpoints, and on
whether $u(x_0)$ vanishes or not (and, if it vanishes, of what
order is the zero), but this is not important for our purposes.
What is important is that when we make the same analysis for
$\sunalldiagramstracenoUsym=b_{LL}$, the only difference (apart
from a factor of $N^{\chi_0}$) is the appearance of the
$N$-independent function $g_{\chi_0}(\beta,\gamma,x)$ in the
integrand. When we perform the $\beta$ and $\gamma$ integrations
the result will just have a factor of $g_{\chi_0}(x) \equiv
g_{\chi_0}(\beta_0(\gamma_0(x),x),\gamma_0(x),x)$ in front of it. So, we get
\eqn\gista{\eqalign{
\sunalldiagramsnoUsym = a_{LL} &\simeq
\frac{(N+1)!}{(N-L)!N^{L}} \int_0^1 dx u(x) e^{Nv(x)}, \cr
\sunalldiagramstracenoUsym =b_{LL} &\simeq N^{\chi_0}
\frac{(N+1)!}{(N-L)!N^{L}} \int_0^1 dx
g_{\chi_0}(x) u(x) e^{Nv(x)} \cr
& \simeq N^{\chi_0}
\frac{(N+1)!}{(N-L)!N^{L}} g_{\chi_0}(x_0) \int_0^1 dx
u(x) e^{Nv(x)}, \cr}}
where the last equality comes from the fact that the
only contribution to the integral comes from the very narrow
neighborhood around $x_0$ \foot{
Here we are assuming that $g_{\chi_0}(x_0)\neq0$. We see no reason why this
should not hold for a generic correlation function and for
generic 't Hooft coupling. When $g_{\chi_0}(x)$ does have
a zero of order $f$ at $x_0$, the contribution of this topology to the
ratio we are interested in will be of order $N^{\chi_0-f}$ instead of
$N^{\chi_0}$, and the leading term of order $N^{\chi_0}$ vanishes.}.

For the ratio we are interested in we simply get
\eqn\ratiosun{
\frac{\sunalldiagramstracenoUsym}{\sunalldiagramsnoUsym} 
\simeq g_{\chi_0}(x_0) N^{\chi_0},
}
which has precisely the expected dependence on $N$. Thus, up to some
assumptions about the existence of appropriate contours, we have
shown that in any $SU(N)$ gauge theory \predict\ has a topological
expansion involving both closed and open string worldsheets.

We have not been able to rigorously justify the contour integrals
performed above, but appropriate contours exist in all examples which
we explicitly checked. There is a special case where the contour
integrations simplify, which is when the theory has a $U(1)$ symmetry
taking $X\rightarrow e^{i\alpha}X$ (this is true in the $\cn=4$ SYM theory).  
In such theories, each component
with no insertions must have $k_1=k_2$, as in
the $SO(2N)$ case. All the formulas can then be rewritten as a single
integral, as in section 4, and we find the same result \ratiosun.  In
this special case, after performing the integration over $\beta$, the
contour integral over $\gamma$ in \betaintegration\ is actually
trivial (the $\gamma$ dependence is exactly $1/\gamma$) so we do not
need to do a saddle-point integral for $\gamma$, but in generic
theories we expect that both saddle point integrations will be
required.

\bigskip

{\centerline {\bf Acknowledgements}}

We would like to thank V. Balasubramanian, S. Minwalla and A. Naqvi
for useful discussions.
This work was supported in part by the Israel-U.S. Binational Science
Foundation, by the IRF Centers of
Excellence program, by the European RTN network HPRN-CT-2000-00122,
and by the Minerva foundation. O.A. would like to thank the Institut d'\'Etudes
Scientifiques de Cargese, the Amsterdam summer workshop,
the Aspen Center for Physics,
the University of British Columbia, and Harvard University for hospitality
during the course of this work.

\listrefs

\end